# Globular Clusters in the Magellanic Clouds – II. IR–array Photometry for 12 Globulars and Contributions to the Integrated Cluster Light

F.R. Ferraro[1], F. Fusi Pecci[1]
V. Testa[2], L. Greggio[2]
C.E. Corsi[3], R. Buonanno[3]
D.M. Terndrup[4], and H. Zinnecker[5]

[1] Osservatorio Astronomico, Via Zamboni 33, I - 40126, Bologna, Italy
[2] Dipartimento di Astronomia, Bologna, Italy
[3] Osservatorio Astronomico di Monte Mario, Roma, Italy
[4] Department of Astronomy, The Ohio State University, 174 W. 18th Ave., Columbus, OH 43210 USA
[5] University of Würzburg, Germany

Send proofs to: F. Fusi Pecci
Osservatorio Astronomico
Via Zamboni 33
I - 40126 BOLOGNA
ITALY
37907::flavio
flavio@alma02.astro.it
fax: 39-51-259407

**Summary**

We report $JHK$ the results of the observations of 12 globular clusters in the Large Magellanic Cloud (LMC), and present the colour - magnitude diagrams down to $K = 16$ (corresponding to $M_K \sim -2.6$) for $\sim$ 450 stars in these clusters. We merge our data with $BV$ photometry for 11 LMC clusters, previously published in Paper I of this series, and use the merged data to study the evolution of integrated magnitudes and colours of *simple stellar populations* (SSPs), which are samples of coeval and chemically homogeneous stars. In particular we examine the effect of *phase transitions* (ph-ts ), which signal the appearance of the RGB or AGB in SSPs of increasing age. We find that AGB contributes $\sim 60\%$ of the integrated cluster light at $K$, while the contribution from the bright RGB stars (*i.e.,* $K_0 < 14.3$, Log $L/L_\odot \sim 2.66$) is correlated with the $s-$parameter (Elson and Fall 1985) ranging from $\sim 0\%$ for $s = 0$ up to $\sim 20\%$ for $s > 35$. The age at which the RGB ph-t actually takes place (i.e., the calibration of $s$ with age) depends on the details of stellar evolutionary models. In "classical" models (those without overshooting), the RGB ph-t occurs at $\sim 6 \pm 2 \times 10^8$ yr and lasts for $2.9 \times 10^8$ yr. In models with overshooting, the occurrence of the RGB ph-t is later (at $\sim 1.5 \pm 0.3 \times 10^9$) and the duration is longer ($4.3 \times 10^8$ yr). While the age and duration of the RGB ph-t depends on the treatment of mixing, both classical and overshooting models yield the same fractional contribution of RGB stars to the total integrated cluster light before and after the RGB ph-t , in agreement with the Fuel Consumption Theorem (RB86). We report extensive experiments which show that the variations of the integrated colours of the LMC clusters from $s = 31$ to $s = 43$ are controlled by the complex interplay of various factors, different from colour to colour and frequently dominated by the stochastic noise induced by few very bright objects. The overall picture emerging is consistent with the early conclusions drawn by PACFM83 and FMB90 that the $J - K$ colour is mostly driven by the AGB stars, $V - K$ is substantially controlled by AGB and RGB stars (AGB stars being slightly more important), $B - V$ is partially influenced by the whole population of red stars brighter than the bulk of the RGB-clump, but it is also quite strongly dependent on the progressive fading and reddening of the turnoff stars due to age increase.

## 1. INTRODUCTION

The globular cluster system of the Magellanic Clouds (MC) provides us with a unique opportunity to investigate the integrated photometric and spectral behaviour of stellar populations as a function of both age and chemical composition. It has been known for a long time (e.g., Baade 1951) that the MC clusters are different in many respects from those in the Milky Way. In particular, they have a wide spread in integrated colours, age, and metallicity (Gascoigne & Kron 1952; Gascoigne 1971, 1980; Danziger 1973; Searle, Wilkinson, & Bagnuolo 1980 – hereafter SWB; van den Bergh 1981 – hereafter vdb81; Hodge 1983; Persson et al. 1983 – hereafter PACFM83; Elson & Fall 1985, 1988 – hereafter EF85, EF88; Bica, Dottori & Pastoriza 1986, – hereafter BDP86; Mateo 1987; Bica, Alloin & Santos 1990; Barbero et al. 1990; Meurer, Cacciari, & Freeman 1990; Frogel, Mould & Blanco 1990, – hereafter FMB90; Seggewiss & Richter 1989; Bica et al. 1991, – hereafter BCDSP91; Bica, Claria & Dottori 1992, – hereafter BCD92).

Because of the wide range in properties that they exhibit, the MC clusters represent the ideal templates to study the evolution of *simple stellar populations* (SSPs), which are samples of coeval and chemically homogeneous stars. Understanding SSPs is vital for interpretations of the evolution of the stellar populations in galaxies over cosmological times. (Renzini 1981, 1991, 1992; Wyse, 1985; Chiosi et al. 1986, 1988, –CBB88; Renzini and Buzzoni 1986, –RB86; Chokshi and Wright 1987; Mateo 1989; Arimoto and Bica 1989; Brocato et al. 1989; Battinelli and Capuzzo Dolcetta 1989; Chambers and Charlot 1990; Barbaro and Olivi 1991; Charlot and Bruzual 1991; Bruzual and Charlot 1993; Girardi and Bica 1992, –GB92; Bressan, Chiosi and Fagotto 1993, –BCF93). Observations of SSPs provide a detailed check of stellar evolutionary models throughout the evolutionary stages experienced by the member stars of the SSP. Understanding SSPs can furthermore lead to the interpretation and precise age-calibration of changes in the integrated magnitudes and colours of the sampled population.

We have already summarized the rationale and specific aims of our project in our previous publications (Renzini 1981, 1991, 1992, RB86, Greggio 1987, Corsi and Testa 1992, Ferraro, Testa, Fusi Pecci, 1993) and in a companion paper dealing with BV CCD-observations of essentially the same MC clusters we will discuss below (Corsi et al. 1993, hereafter *Paper I* ). We will therefore briefly elaborate on the characteristics of SSPs before presenting our infrared photometry.

Stellar evolution theory (see for example RB86) predicts that red stars dominate the bolometric luminosity of an SSP after its first evolutionary stages. As is well known, the main red features of the CMDs in clusters and other SSPs are the Asymptotic Giant Branch (AGB) and the Red Giant Branch (RGB). Their extent and contribution to integrated magnitudes and colours depend on the age and metallicity of the stellar population. According to the classical picture, the AGB appears quite abruptly when the first stars develop a degenerate C-O core. In standard models (*i.e.*, without overshooting, see Sweigart et al. 1989,1990, Castellani, Chieffi and Straniero 1992, and Chiosi, Bertelli, and Bressan 1993 for discussions and references) this occurs when stars less massive than $\sim 5M_{\odot}$ evolve off the main sequence, at a cluster age of roughly $10^8$ years. Similarly, the extended RGB appears when the evolving stars develop a degenerate He-core, which takes place when stars less massive than $\sim 2.2M_{\odot}$ evolve off the main sequence, corresponding to a cluster of roughly $6 \times 10^8$ years (Sweigart, Greggio and Renzini 1989, 1990). Models taking into account a mild overshoot (BCF93) behave similarly, the only difference being in the lifetimes of core H- and He-burning phases and in the mass range limits. In particular, according to BCF93, the age corresponding to the appearance in the CMD of AGB objects increases by about a factor 2, while that associated to the full development of the RGB is about 20% larger. Within this framework, one could therefore predict that rapid variations in the integrated magnitudes and colours of the SSP, called *Phase Transitions* (ph-ts ) after Renzini and Buzzoni (1983, 1986), would occur at known ages as a consequence of the appearance in the CMD of AGB or RGB bright stars. In this respect we notice however that recent evolutionary computations by Blocker and Schönberner (1991) indicate that the appearance of an extended, well populated AGB may well be delayed with respect to the previously quoted ages (Renzini 1992). It has been found that AGB stars experiencing the envelope burning process climb quickly to very high luminosities, where they are likely to suffer severe mass loss, thus leaving soon the AGB. This effect then leads to a substantial shortening of the lifetime of the more massive stars in the bright portion of the AGB, leaving instead unaffected the evolution of those stars whose mass is too low to experience the envelope burning process. As a result, the age of the AGB phase transition gets closer to that of the RGB phase transition. Renzini 1981 and RB86 advanced the hypothesis that integrated populations would experience observable ph-ts which could be dated via understanding of SSPs, and described the possible use of the ph-ts as powerful indicators of galaxy ages. Their original hypothesis has subsequently been studied and questioned by various authors (see especially BCF93 and references therein). In particular, based on their model computations, BCF93 conclude that the ph-ts cannot be used as age indicators when evolutionary and cosmological effects are fully taken into account.

On the other hand, BCF93 do not deny that the ph-ts (i.e. the "sudden" appearance of AGB/RGB red stars in a SSP) actually take place, but instead show that in a galaxy the predicted ph-ts are masked due to a combination of various factors. Nevertheless, a comparison of SSPs with stellar evolution theory will be of importance, since such a comparison permits the measurement of the contributions to the integrated cluster light of the individual evolutionary phases. Since the intermediate-age MC clusters are presumably the best available SSP's covering such an interesting range in ages, they are therefore the best tool to use to verify the actual existence of the ph-ts , to identify precisely at which age AGB and RGB stars first appear in an evolving SSP and, finally, to evaluate the specific contributions of these objects to the integrated cluster light.

Intermediate-age clusters in the MCs have intermediate colours ($0.3 < (B-V) < 0.6$). They also have type IV-VI in the SWB-type classification, and have $s = 30 - 45$ in the classification of EF85 and EF88. CMDs from RB86 and *Paper I* reveal that the SWB-type III clusters or earlier (blue or young clusters) do not display an extended AGB or RGB sequence, while most of the SWB-type V objects or later (red or old clusters) have a well populated AGB and RGB. The SWB-type IV represents thus the "transition" class where major integrated color variations occur.

Further evidence of the peculiar importance of the study of these transition clusters emerges from the analysis of the distribution of the integrated MC cluster colours plotted in Fig. $1a - d$ versus the parameter $s$ defined by EF85. The data have been taken from EF85: $s-$values, vdB81: B-V colours, PACFM83: $V-K$, $J-K$, $H-K$. In the diagrams, the clusters chosen for our survey are marked as full dots and identified with their NGC number. It is quite evident from the plots that the clusters here considered are located (especially in the $B-V$ and $V-K$ diagrams) where significant variations in the integrated colours take place. The near-IR CMDs offer thus the best possibility to study the red (and rather cool) stellar sequences, and therefore to identify the specific integrated magnitude and colour glitch, if any, their appearance could originate.

The present paper presents part of the results of a pilot project started in 1985 on a photometric (BVJHK) study of a sample of intermediate-age clusters in the MC.

As already stated, BV CCD-photometry has been reported in the companion *Paper I* . Besides presenting the photometric data, we also add and discuss here the results obtained by carrying out a series of experiments and simulations based on the whole set of BVJHK data.

## 2. THE CLUSTER SAMPLE

The list of the 12 MC clusters for which we have obtained infrared photometry is presented in Table 1. Additional data, mostly from optical photometry, can be found in Tables 3 and 4 of *Paper I* . In Table 1 are displayed *col. 1*: the NGC number of each cluster, *col. 2*: the integrated K magnitude, *col. 3,4,5*: the integrated intrinsic $V-K$, $J-K$, $H-K$ colours from PACFM83, *col. 6*: the integrated $B-V$ colour from vdB81, *col. 7*: various estimates of the individual cluster reddening, *col. 8*: the SWB-type, *col. 9*: the value of the $s-$parameter as defined and measured by EF85.

Most of the clusters in this paper were discussed in *Paper I* , and general information on individual clusters can be found there (see Tables 3,4). There are three clusters which were not included in *Paper I* which have optical photometry in the literature:

$\bigoplus NGC1783$

NGC 1783 was classified as "old" by vdB81 and SWB-type V by SWB. EF85 and EF88 determined $s=37$, later corrected for reddening using UV-colours by Meurer et al. (1990) to $s_0=38$. Several estimates of age and metallicity are available from various authors, using different approaches (BDP86, using spectral features, Mould and Aaronson, 1980 (AMMAI), Aaronson and Mould, 1982(AMMAII), using AGB stars in the IR bands, Mould et al. , 1989, using optical bands) as summarized in Tables 3,4 of *Paper I* . The late-type stellar content of NGC 1783 was extensively studied by Frogel and Cohen (1982), PACFM83, Cohen 1982, AMMAI,II. Specific considerations on its AGB were made by FMB90.

Previous CMDs are available from Sandage and Eggen (1960), Gascoigne (1962), and Mould et al. (1989). Structural and kinematical parameters were also determined by Freeman et al. (1983), Kontizas et al. (1987), and Olszewski et al. , 1991 (OSSH91).

$\bigoplus NGC1806$

NGC 1806 was included by vdB81 in his list of "old" clusters. It has SWB-type V and $s=40$ in the classifications by SWB and EF85, EF88, respectively. Age estimates are available from BDP86 and AMMAII (see Table 4 *Paper I* ). Detailed IR studies were made or revised by FMB90. UV magnitudes and colours were secured and discussed by Barbero et al. (1990). Freeman et al. (1983) and Kontizas et al. (1987) give structural and kinematical parameters.

An optical CMD was presented by Geyer and Hopp (1982), but we were unable to reliably identify the objects in common. Hence, we could not compute $V-K$ colours for any star in this cluster.

$\bigoplus$ $NGC1978$

NGC 1978 is "old" in the classification by vdB81 and has SWB-type VI and $s = 45$ in SWB and EF85, EF88, respectively. In their revision of the $s-$values, Meurer et al. (1990) define the reddening-free parameter $s_0$ and give for NGC 1978 $s_0 = 43.8$. This cluster has been so far the subject of various studies, including a recent paper by Fischer, Welch, and Mateo (1992) on its dynamical properties. Age and metallicity estimates are available from various authors (see Table 4 in *Paper I* ). In particular, Chiosi et al. (1986) presented various age estimates using different approaches and different models (classical and with overshoot). As for NGC 1783, Mateo (1992) made an analysis of the temporal evolution of the integrated $M_V$ showing that this cluster will probably fade from $M_V \sim -8.5$ to $M_V \sim -7.2$ at age $t = 15 Gyr$.

Optical data were published by Olszewski (1984) but we were unable to firmly identify a set of stars in common.

## 3. OBSERVATIONS AND REDUCTIONS

### 3.1 Observations

Infrared images of 12 MC clusters were obtained with the 1.5 m telescope of the Cerro Tololo Inter-American Observatory (CTIO) on November 27-30, 1988. The detector was the IRIM camera (NOAO Newsletter, March 1987), which is based on an InSb IR-array, 58x62 $px$ , with a pixel size of 0.75 $\mu = 0.92''$, and consequently a field coverage of about $1' \times 1'$. Each cluster was mapped by 4 partially overlapping fields to cover a squared area, roughly $2' \times 2'$, centered on the cluster center. We also typically obtained several "sky" frames near the target clusters. The use of these frames in the reduction is described below.

During the run the weather conditions were stable with an average seeing of $\sim 1.2''$ FWHM. The total integration time per field was typically 2–3 minutes in each of the $JHK$ bands; the full integration times were obtained by co-adding short exposures, typically with integration times of 45 sec.

### 3.2 Reductions

#### 3.2.1. *Linearity and corrections to uniform sensitivity*

Initial data reduction (de-biassing, linearization and sky correction) was performed using MIDAS, the standard ESO reduction package. A complete description of the procedures necessary to properly calibrate the frames can be found in the review by McCaughrean (1989).

The first step in the processing of the IRIM frames was a correction for small non-linearities in the response of the InSb detector. The correction included terms up to quadratic in intensity, but was small ($< 2\%$) over the intensities of the stars in the target clusters. The next step was to subtract a zero-exposure, or "bias" frame, from all the data frames. The bias frame was generated from the average of a large number of frames with the shortest available exposure time (0.075 sec).

Then from each frame was subtracted a "dark" frame of equal exposure time to the cluster or sky frames. The dark frame was constructed from the average of a large number of frames obtained with a cold stop in the light path. Over most of the detector, the dark current was insignificant in comparison to the brightness of the sky, but there were several regions of the chip which produced a large dark current. The next process was to combine the several bias-subtracted, linearized sky frames obtained near each cluster into a mean image which measures the average sky brightness at the time of the cluster observation. These were multiplicatively scaled to a common mean, then merged with a median operation to remove any stellar images on the sky frames. The bad pixels on the merged sky frames were then interpolated over, and the sky frames were subtracted from each adjacent cluster frame.

The final process was to construct a superflat image, which was composed of the mean of all sky frames on a given night, after these had been scaled to a mean of unity. The superflat image, which had high s/n, was then divided into the sky-subtracted frames of each cluster.

#### 3.2.2. *Photometry, calibration and photometric errors.*

Seven standards from the list of Elias et al. (1982) were observed to calibrate the instrumental IR data onto the standard CTIO/CIT system. Instrumental magnitudes were obtained using a synthetic aperture of diameter $\sim$ 30 pixels. Now due to a problem with a capacitor improperly installed in the read-out electronics, stellar images contained an extended tail in the readout direction of the detector. For the brightest stars, $\sim$ 83% of the light fell within an aperture of 3.6" and $\sim$ 94% within 7.2", while the residual part was dispersed in a tail reaching up to 10-20 *px* or even farther. We assumed that the fraction of light in the tail was independent of the brightness of the stellar image.

Each standard was repeatedly observed during each night; the r.m.s. scatter on a given night was always less than 0.02 mag. Five standards were observed on two different nights, yielding an r.m.s. scatter for each which was less than 0.03 mag.

We derived the transformations between the instrumental magnitudes and the CTIO/CIT values to be

$$J = j + (18.838 \pm 0.047) - (0.145 \pm 0.020)(j - k)$$
$$H = h + (18.102 \pm 0.042)$$
$$K = k + (17.827 \pm 0.026)$$

where $j, h$ and $k$ are the instrumental magnitudes and the errors are the statistical errors in the coefficients. Residuals before transformation to the standard system are displayed in Fig. 2. We did not find evidence for a colour term in the $H$ and $K$ bands, but our photometry did not include a significant spread in colour of the standard stars. Photometry of the cluster stars was carried out using a version of the package ROMAFOT (Buonanno et al. , 1979,1983) specifically optimized for the treatment of undersampled images (Buonanno and Iannicola 1989) and mounted on a DEC-station 5000/240 at the Osservatorio Astronomico di Bologna. ROMAFOT characterizes the undersampling by the parameter $R = F/P$, where $F$ is the FWHM of the stellar images and $P$ is the size of a pixel. Within ROMAFOT, the indicative limiting value $R_{lim}$ separating the undersampling and oversampling regime is $R_{lim} \sim 1.7$ (see Buonanno and Iannicola 1989). Typically the stellar images on our frames were strongly undersampled ($R \sim 1.3$), so the use of a package such as ROMAFOT was critical.

Object detection was carried out independently in each field using the standard procedure available in ROMAFOT and already described briefly in other papers (see Ferraro et al. 1990). Because faint stars could not be reliably detected in the extended tails of brigher objects, we decided not to push the detection algorithm to the faintest possible limit.

Relative photometry for the stars on each frame was measured using a two-dimensional fitting procedure. The transformation of the magnitudes from ROMAFOT to the instrumental system was performed by measuring aperture magnitudes in the same manner as for the standards on the most isolated stars on each frame. This step introduced an additional error on the photometry since it was not always possible to determine the transformation to aperture magnitudes accurately, given the small size of the IR-array, the presence of a noisy background, and extended image tails. In order to estimate the total photomeric errors, one should in general consider the combination of different sources. Internal uncertainties can be determined from multiple measurements of stars falling in the frames' overlapping regions. Following this approach, we estimate that the total internal photometric accuracy of our measurements is about 0.05 mag for the bright stars and about 0.1 mag for the faintest ones. Then, considering the errors introduced by the procedure adopted to match the aperture-magnitudes to the profile-fitting-magnitudes and those possibly affecting the zero-points, we estimate a conservative total uncertainty of about 0.15 mag.

#### **3.2.3.** *IR Magnitudes and Colours and Comparison to Previous Photometry*

The final adopted IR magnitudes and colours are presented in Table 2, the first column of which lists an identifying name for each star, followed in the next three columns by the $K$ magnitude and colours. Columns 4 and 5 of Table 2 display the $V-K$ and $B-V$ colours for those stars with reliable cross-identification with the photometry in *Paper I* . The final two columns give the position of each star expressed in pixels, at a scale of 0.92 arcsec pixel$^{-1}$. Finding charts for the stars in each cluster are shown in Fig. $3a-l$; the $x$ and $y$ positions on the charts are as in Table 2.

Although we took special care in identifying the stars measured here with those listed in previous work, the total number of stars in common is quite small, for several reasons. First, we have observed central regions of each cluster, while most previous surveys were with aperture photometry in the outskirts of the clusters. Second, though some identifications were made possible by the use of a set of original maps kindly made available to us by Dr. J. Mould, a few stars located in very crowded regions could not be positively identified. Finally, our survey is deeper than reached before and many stars have no earlier counterpart simply because they are fainter than the limit of previous photometry.

The IR magnitudes and colours for the brightest stars of the present photometry can be compared with those reported by AMMAII and FMB90. Based on the stars identified in Table 3, the residuals (in the sense *others-this paper*) are plotted in Fig. $4a - c$ versus our magnitudes. In the figure different symbols refer to different clusters.

As can be seen, the scatter of the various points is quite high but always less than 0.3 mag. Concerning the zero-point, it is worth noting the fact that while in the K and H bands the residuals do not show any significant systematic shift compared to the previous photometries, a systematic difference (maybe colour dependent) of $\sim 0.10$ mag is evident in the J-band, with our photometry being fainter than the previous ones. We do not have any ready explanation for such an effect since the quoted possible causes of errors in our procedure should have a very similar impact on the three used bands. On the other hand, we have not found any indication from our data to arbitrarily shift our magnitudes to those of the previous systems. It is thus clear that, were our J-magnitudes to be revised by such a zero-point shift, also the colours involving J-magnitudes should be accordingly corrected. A further check has been made to verify the existence of any trend in the residuals (our data minus previous photometry) as a function of the crowding conditions. Clusters have been ranked into three classes of crowding obtained by a simple inspection of the frames. We have thus classified NGC 1783, 1806, 1978 as "severely crowded", NGC 1756, 2162 as "medium crowded" and NGC 1831, 1987, 2173, 2108, 2209 as "not crowded", with the caveat that NGC 1987 and 2108 lie in a very field-contamined area (the LMC Bar). We have then computed the residuals for each cluster, obtaining that a small trend seems to exist in the direction of making the differences *others - this paper* more negative with increasing crowding, with the exception of NGC 1978 that has very small differences in the photometry (filled circles on Fig. 4). It seems therefore plausible that the previous aperture photometry in the most crowded clusters maybe affected by the inclusion in the diaphragm of small surrounding stars. However it is important to stress here that, since most of the results presented and discussed in the following sections are essentially based on overall properties of groups of stars and on "relative" quantities, the quoted possible difference in the zero-point of the J-band should not affect significantly our main conclusions.

## 4. THE COLOUR-MAGNITUDE AND COLOUR-COLOUR DIAGRAMS

### 4.1. The K, J-K CMDs and a first rough separation into two main groups

The final CMDs for each cluster are plotted in Fig. 5. They include all the stars listed in Table 2 and represent the basic sample used in the following analysis and discussions. As can be seen from the various diagrams, our samples reach $K_0 = 16$, where $K_0$ is the dereddened K magnitude, but because the photometry is very uncertain at faint magnitudes, we have limited our analysis to $K_0 < 14.3$. From a first inspection of the CMDs, we find:

1. Most of the CMDs contain a clump of faint stars usually spread out in colour, and a giant branch which has a similar slope in all clusters for which a ridge line can be drawn.
2. The brightest star is usually brighter than $K = 11$ but it is generally not easy to measure the brightness of the tip of the giant branch. Not only is the membership of the brightest star in the cluster uncertain, but there are usually a small number of stars at the top of the giant branch.
3. Although some CMDs contain only a few stars, it seems evident that the bright giant branch is well populated in some clusters and almost totally lacking in others.

Previous authors (Lloyd Evans 1980, Mould and Aaronson 1979, Frogel et al. 1980 (F-PC80), AMMAI, AMMAII, PACFM83, Frogel 1984,1988, FMB90) have reached similar conclusions based on aperture photometry.

Based on inspection of the CMDs, we have divided the clusters into two groups. The first group includes those clusters with a populated bright giant branch, while the second contains those where the giant branch is weak or absent. The first group includes NGC 1783, 1806, 1978, 1987, and 2173. The clusters NGC 1756, 2107, and 2209 are in the second group. The membership of the others is uncertain.

This simple division of the clusters by giant branch morphology correlates with the SWB-types (*see* Table 1). The clusters inserted in the first group have (see Table 1) mostly SWB-type V-VI and $s \sim 35 - 45$, while those in the second one have SWB-type III-IV and $s \sim 30 - 35$. If NGC 1987 is excluded from the first group (as shown in *Paper I* , it is severly contaminated by field stars even in its central regions), the first group would contain clusters classified no earlier than SWB-type V and $s > 37$, and would be fully separated from the second group in SWB type and $s$ parameter.

In the light of the discussion of SSPs above, we present the following interpretion of the two groups of clusters:

a) The clusters without a well developed giant branch are those where the masses of the presently evolving stars are still larger than the critical mass to produce classical AGB or RGB stars. The clusters of the second group are older and, consequently the evolving stars have masses appropriate to yield numerous AGB and/or RGB stars. The clusters which could not be divided into the two groups are intermediate-age clusters, at the transition between the older clusters with giant branches and the younger clusters which lack them.

b) Because the giant branch stars dominate the integrated red or infrared light of the clusters, the presence or absence of a giant branch would explain the observed dependence of integrated colour on SWB-type or the $s-$values, at least for $B-V$ and $V-K$: the clusters with a well populated giant branch have redder intrinsic colours. The behaviour in the $J-K$ and $H-K$ colours is more complex and is highly dominated by statistical fluctuations and the incidence of a few bright (carbon) stars (FMB90).

To further illustrate the differences between the two cluster groups, we assumed that the separation of the two groups of clusters occurs at $s = 35$ (see *Paper I* ), then constructed composite CMDs for the clusters in each group. These CMDs are shown in Fig. 6, where the upper panel shows all the stars from the second group, and the lower panel contains the stars from the group with giant branches. Dereddened magnitudes and colours have been obtained assuming $E(B-V)$ of the various clusters listed in Table 1 and the extinction curve of Savage and Mathis (1979) which gives $A_K/E(B-V) = 0.38$ and $A_J/E(B-V) = 0.87$. We divided the CMDs into three intervals of $K_0$ as shown by the dashed lines in Fig. 6, and computed the relative numbers of stars in the three magnitude intervals. The three magnitude intervals are: $K_0 < 12.3$ (all colours), $12.3 \leq K_0 \leq 14.3$ with $0.4 \leq (J-K)_0 \leq 1.2$, and $14.3 < K_0$ (all colours). The colour selection for the middle magnitude group was made to correct for field contaminants (see the discussion below). For the clusters without strong giant branches, the fraction of stars in each magnitude bin is (0.10:0.13:0.77), while the distribution for the clusters with giant branches is (0.12:0.39:0.49). From this we conclude that the principal difference in the integrated light of the clusters between the two groups is that the clusters with giant branches have a greater contribution from stars at intermediate magnitudes, rather than having a different fraction of stars at high luminosity.

Although plausible to first order, these conclusions must be scrutinised by a deeper analysis. For example, it is clear from Fig. 6 itself that the simple subdivision of the clusters in two main groups is not a clean division; in particular there are a considerable number of bright stars even on the clusters which were in the group with weak giant branches. In the next several sections, we discuss the CMDs in greater detail.

### 4.2. Background contamination

The CMDs in this investigation do not provide much information about contamination by stars in our galaxy or in the field of the MCs, for the simple reasons that we observed only the central regions of the clusters and did not obtain field CMDs away from the clusters, so that it is quite possible that some stars we identify as giant-branch members of the clusters are in fact background stars. On the other hand, the fact that we observed only the central regions of each cluster reduces (we hope) the importance of field contamination. As a comparison, FMB90 adopted as cluster members almost all the stars inside a circle having diameter $1'$.

Better information on the field contamination may be found in the optical study we presented in *Paper I* , which was based on CCD frames covering a much larger area than do the infrared frames here. From the surface density of stars in radial zones centered on the clusters, we conclude that only NGC 1987 has a strong field contamination in the area surveyed with the IR-array. In the following discussion, we will assume that most of the stars on our CMDs are in fact members of the cluster, with the exception of those stars far from the ridge line of the mean giant branch in Fig. 6. We therefore have excluded stars in the range $12.3 < K_0 < 14.3$ that have colours $(J-K)_0 > 1.2$ or $(J-K)_0 < 0.4$ from our analysis.

### 4.3. The two-colour diagram

The near-IR two-colour diagram for all the stars in Table 2 is shown in Fig. 7. The composite plot naturally has a stronger statistical significance than those for the individual clusters; the individual two-colour plots for each cluster can easily be derived from the data in Table 2. The regions of the two-colour diagram which are occupied by carbon stars and LPVs are also marked on Fig. 7). The limits for these regions are taken from Bessell and Brett (1988), Frogel and Elias 1988, AMMAII, Cohen et al. 1981 (CFPE81), Frogel et al. 1978 (FPAM78). Many authors have previously discussed the use of the two-colour diagram to distinguish various types of stars (especially see FMB90 and references therein), so in this discussion we will be very brief.

There is considerable scatter in the two-colour diagram. Most of the scatter for the faint stars (those with bluer colours in Fig. 7) can probably be ascribed to photometric errors. On the other hand, these stars have been measured in the very central regions of the clusters and such a spread (especially amongst the fainter objects) may be caused by crowding. Due to the residual uncertainty in the J-band zero-point, there might also be a residual systematic vertical shift.

From Fig. 7 and the previous figures, it is evident that most of the stars in our sample are K and M giants, but there are also stars with colours typical of carbon stars and LPVs. These stars have all been detected in previous investigations since they are the brightest objects in each cluster; we therefore will not analyse them further. The two groups of clusters (above) behave quite differently in the near-IR two-colour plane. Fig. 8 displays the two-colour diagram for (upper panel) all the stars in the first group of clusters, and (lower panel) those in the second group. Although the total number of stars is quite different in the two samples shown in Fig. 8, both the Carbon star and LPV regions are actually unpopulated in the "early-type" clusters (panel *a*). Two stars fall in the Carbon/LPV stars area: star # 3 of NGC 2209 (see PACFM83, FMB90) and # 26 of NGC 1831 (AMMAII).

### 4.4. The K,V-K CMD and the separation of AGB stars

As repeatedly remarked here and by previous studies, the two most important contributors to the cluster integrated light are the RGB and AGB stars. Though in the best CMDs of Galactic globular clusters AGB stars can frequently be separated from RGB members almost up to the giant branch tip, such a subdivision is impossible in the MC clusters as these two types of stars are not easily distinguishable due to photometric errors in the region of the CMD where they partially overlap. It is nevertheless important to attempt a separation of the RGB and AGB stars, since in a SSP their first appearance in the observed CMD occurs for different evolving stellar masses and, thus, at different cluster ages.

The specific problem of the separation of AGB and RGB objects in the MC clusters has already been deeply discussed by FMB90 (and references therein). We will adopt their procedure, which is based on the choice that all the stars brighter than $M_{bol} = -3.6$ are exclusively AGB stars, whilst the fainter ones are mostly RGB objects.

This choice is in perfect agreement with SGR90 models which yield $M_{bol}$ = -3.57 for the RGB tip of models of the appropriate chemical composition. On the other hand, some of the brightest stars fainter than $M_{bol}$ ∼= -3.6 could be Early AGB (E-AGB) stars. Stellar models spend ∼ 14 Myr on the E-AGB (Renzini and Fusi Pecci 1988), and ∼ 50 Myr on the RGB, at luminosities brighter than the Helium burning clump (SGR89). Thus the fraction of E-AGB stars at $M_{bol}$ fainter than -3.6 can be estimated to be around 20%, and this fact has to be taken into account when computing the contribution of just the RGB stars.

To determine the K-magnitude at which the actual separation between AGB and RGB stars could be set in our samples, we have adopted a corrected distance modulus $(m - M)_o = 18.6$, (see Westerlund 1990, for a review). We adopted a bolometric correction, $BC_K$ =2.62 at $(V-K)_0$ =3.8 based on the $K_0, (V - K)_0$ CMD presented in Fig. 9 (including the stars measured in the various clusters for which the identification on the corresponding V-frames was feasible) and the calibration presented in Fig. 11 of FPC91 for $BC_K$ vs $(V - K)_0$. With these assumptions we eventually located the separation threshold between AGB and RGB stars at $K_0 = 12.3$. We have measured the mean RGB ridge line in the $K$, $V - K$ diagram and have used this to derive an estimate of the average cluster metallicity. The determination of the ridge line is relatively secure because the stars with a reliable identification in the optical and IR samples are generally those with the most accurate photometry. We find $(V-K)_{o,GB} = 3.3$ at $M_{K_o} = -5.5$. Via the calibration against [Fe/H] presented by Frogel et al. (1983 – FPC83), we get [Fe/H]=-1.56 (*i.e.* Z = 0.0005), consistent with other determinations of the mean metallicities of these MC clusters (see OSSH91, and references therein).

Another metallicity estimate can be obtained following the approach of Davidge et al. (1992). We have thus measured the $\Delta(V - K)$ as the difference in colour between the points on the GB with $M_K \sim -3$ and $M_K \sim -5$. In our sample, $\Delta(V - K) \sim 0.7$. Using the calibration against [Fe/H] in their Fig. 6 for nine Galactic globular clusters, we obtain [Fe/H]$\sim -1.2$ for our sample. Taking into account the uncertainties affecting both methods ($\pm 0.2$ *dex*) the two figures are compatible within the errors.

## 5. THE RGB PHASE TRANSITION AND CONTRIBUTIONS TO THE INTEGRATED CLUSTER LIGHT

In this section we report the results of some simple calculations which estimate the fraction of light in various bands that is produced by stars in the RGB and AGB phases of evolution. These calculations are necessarily unsophisticated, since the sample of clusters studied here is rather limited and because the available star sample in each cluster is small and possibly incomplete. Therefore, it may be risky to use the results as a firm basis for yielding a *quantitative* analysis of the overall photometric properties of the clusters and of the specific contributions of the various evolutionary phases to the cluster integrated light.

### 5.1 Normalised numbers of AGB and RGB stars

We define the parameter $N_4$ to be the number of giants within the range $10 < K < 14.3$, normalised to the luminosity of each cluster in units of $10^4 L_\odot$. Table 4 lists the values of $N_4$ for each cluster, while Fig. 10(a) shows the correlation between $N_4$ and cluster type $s$. As expected from evolutionary theory, the number of bright (red) stars per unit luminosity increases with the age of the cluster up to about $s \sim 37$, then possibly levels off for older clusters. This behaviour reflects the development of a populated giant branch, and is consistent with the observation that the clusters of higher $s-$value have redder integrated colours. We now consider the contributions from AGB and RGB stars separately by defining two other parameters: let $N^*_{\rm AGB}$ and $N^*_{\rm RGB}$ be, respectively, the number of AGB and RGB stars normalised to $10^4 L_\odot$. The values of $N^*_{\rm AGB}$ and $N^*_{\rm RGB}$ are listed in Table 4 and plotted in panels (b) and (c) in Fig. 10. In computing these parameters, we have taken into account that about 20% of the stars in the range $12.3 < K_0 < 14.3$ may consist of Early-AGB objects. Since within this magnitude interval we do not know which stars are actually E-AGB members, and since the E-AGB stars are predicted to be in general as bright as the bright portion of the RGB, we have selected a few (usually 2-4 stars in total) randomly among the 10 brightest stars in the considered bin. This may introduce some bias, but, given the small size of the sample, its possible effects should be always within the uncertainties due to the statistical fluctuations.

As can be seen from Fig. 10(b) and (c), the normalised numbers of AGB stars are strongly dominated by the statistical noise intrinsic to the available samples. Looking at the plot (Fig. 10*b*), one could perhaps only conclude that, while AGB stars seem to be infrequent in the early type clusters, their number fluctuates with increasing $s$.

RGB stars are fainter but more numerous, so their normalised numbers are more certain. Interestingly enough, their behaviour is quite similar to the overall cluster distribution in panel (a). Though fainter in general than AGB members, RGB stars drive the overall trends when the analysis is based just on star counts. On the other hand, one has always to recall that, since AGB stars are brighter and redder, just one of them may contribute more to the total cluster integrated light than the whole RGB population. Hence, this analysis has to be repeated using the star light and not just the star numbers.

Before proceeding further, it is interesting to compare the data in Fig.10 with the corresponding theoretical expectations. The number of stars $N_j$ in a post-MS evolutionary phase is proportional to the time $t_j$ spent in that phase according to the relation (RB86) $N_j = B(t) \times L_{tot} \times t_j$, where B(t) is the specific evolutionary flux and $L_{tot}$ is the total luminosity sampled. For ages older than the RGB phase transition B(t) $\sim$ $2 \times 10^{-11}$ stars/yr/$L_{\odot}$ (Guastamacchia 1992). The time spent on the E-AGB at magnitudes brighter than $M_{bol}$ = -10 is $\sim$ 1.7 Myr, and a comparable amount of time is spend during the Thermally Pulsing AGB regime (Iben and Renzini 1983). Model stars spend approximately 7 Myr on the RGB at magnitudes brighter than $M_{bol}$ = -10, and then for every $10^4$ $L_{\odot}$ sampled, we expect to find $\sim$0.7 AGB and $\sim$ 1.4 RGB stars brighter than $M_{bol}$ = -10, which turns out to be close to what we observe.

### 5.2 The overall AGB and RGB contributions to the cluster integrated light

Table 5 lists the AGB and RGB contribution to the luminosity of the cluster in K in columns 2 and 3, respectively. In Fig. 11 we show the plots of the fractional contributions as a function of $s$ for the AGB in Fig. 11(a), and RGB (b), respectively. These contributions are normalised to the luminosity of the cluster as derived from $K$-band aperture photometry. From Fig. 11(a), we conclude that in most clusters the contribution of AGB+RGB stars to the total integrated light is close to 70%, or even larger. There is a weak increasing trend with $s$, though perhaps a result of statistical fluctuations, which is in *qualitative* agreement with the corresponding plot for the normalised counts [Fig. 10(a)]. In this plot there is no trend which shows the existence and impact of the RGB ph-t .

As shown in Fig. 11(b), the AGB contribution is always very high (up to $\sim$ 60% or more), though again strongly fluctuating as expected given the very small numbers of stars involved and their very high intrinsic brightness. Therefore, whenever present, AGB stars dominate the integrated cluster light. Since they are very red and few, they actually drive the IR magnitudes and colours, including their fluctuations. The RGB contribution is in general smaller than the AGB [Fig. 11(b)]. There is a strong increase of the contribution from the RGB stars at values of $s \sim 35$: the mean contribution of the RGB stars for the clusters with $s < 35$ is $0.02 \pm 0.01$ while for those with $s > 35$, excluding NGC 1987 and NGC 2108, the contribution is $0.20 \pm 0.01$ (the two distributions differ at a $5\sigma$ level). As with the star counts, the effect of the RGB ph-t is clearly shown. In order to make more direct the comparison with theoretical models, we have also computed the RGB contribution in bolometric flux (following the assumptions described above). The result is listed in column 5 of Table 5 and is shown in Fig. 12. As already noted in the K band (see Fig. 11a), there is a clear increase of the RGB contribution to the bolometric light for $s > 35$, the mean contribution increasing from $0.007 \pm 0.004$, for clusters with $s < 35$, up to $0.18 \pm 0.01$ for those with $s > 35$ (excluding NGC 1987 and NGC 2108).A Student's $t$-test has been performed in order to check how significant is the difference between the two distributions. The $t$-test variable turns out to be $t = 22.04$, with eight degrees of freedom. Hence, the two distributions differ at a confidence level higher than 99.99%. We also computed the combined r.m.s. using the usual formula $\sigma_{tot}^2 = \sigma_1^2 + \sigma_2^2$. In this picture, the difference of the two average values is about $12\sigma_{tot}$.

### 5.3 Dating of the RGB ph-t and Comparison with Theoretical Models

Up to this point we have reached conclusions which are independent of evolutionary theory, though we point out that our method of analysis was guided by predictions of evolutionary models (for instance in RB86). We have presented above an observational confirmation of the existence of the RGB ph-t , but we have not yet determined the absolute age at which the RGB ph-t takes place, nor have we presented a detailed *quantitative* comparison with the evolution theory. We now proceed to calibrate the age of the RGB ph-t .

Before carrying on, however, we note that reliable age determinations should only be based on fully calibrated theoretical isochrones (see, for example, the discussions in Öpik 1938, Sandage and Schwarzschild 1952, Renzini 1991, Fusi Pecci and Cacciari 1991). Therefore, in principle the best procedure for calibrating the age of the RGB ph-t would be: (i) adopt a well-calibrated set of theoretical isochrones, (ii) adopt a calibration of $s$ with age, and thereby understand the trends with $s$ shown in Figs. 10–12; (iii) determine the age at which the essence of the RGB ph-t become detectable and the time duration necessary to have its completion, *i.e.* determine the age of the cluster where the first RGB stars appear and that of the cluster where the RGB is fully developed and populated.

In practice, however, we are forced to deviate significantly from this ideal procedure and are unable to reach a unique age calibration for the RGB ph-t for the following reasons:

(1) At present, there is no fully tested and calibrated set of theoretical tracks unanimously adopted to accomplish items (i) and (ii) above. In particular, there is still an ongoing discussion about the treatment of mixing phenomena (and especially overshooting) which affect the H-burning and He-burning lifetimes (see BCF93 and references therein). Hence, the adopted time-scale depends on the basic assumptions made in the model computations. Furthermore, the effect of the envelope burning process on the AGB evolution has not been systematically explored yet.

(2) Even if one adopts a set of models as the "correct" reference grid, the comparison between the models and the observations requires the definition of consistent quantities to be determined from the observational data, on the one hand, and from the theoretical tracks, on the other, which guarantee a meaningful test. This is far from trivial.

(3) If one attempts to use real clusters to bracket in time the starting age and the duration of the whole RGB ph-t , one should have a much larger sample of clusters, properly distributed in $s-$parameters and in their sub-classes, and much bigger samples of individual stars measured for any evolutionary stage than we have in this study.
Consequently, we will here discuss the effects of overshooting on the $s$–age relation, define which quantities can be determined from the models to make comparisons with the observational data, and give a range of mean ages and time-intervals concerning the RGB ph-t .

**5.3.1.** *The "s-age" calibrations*

Among the many calibrations of the $s-$parameter in terms of age presented in the literature (see CBB88, EF88, MCF90, BCF93 for references), we have chosen to adopt just two of them, which we will term "classical" and "overshooting" models and which one could derive by estimating the age of the calibrating clusters by using "classical" or "overshooting" models, respectively. In particular, we hereafter call "classical" the calibration

$$\log t = 0.079s + 6.05, \quad (1)$$

obtained by EF88. We will use under the name "overshooting" the calibration recently obtained by Chiosi et al. 1993:

$$\log t = 0.067s + 6.17, s \leq 26, \quad (2a)$$

$$\log t = 0.180s + 3.22, 26 < s \leq 31, \quad (2b)$$

$$\log t = 0.067s + 6.73, s > 31, \quad (2c)$$

and based on the models computed by Alongi et al. (1993–A193). These two calibrations are representative of the several that are available, but note the number and quality of the calibrating clusters differ from author to author. Even with the same clusters, the calibrations depend on the adopted MC distance modulus. Therefore significant differences in the age calibration as a function of $s$ can be found in the literature. For example, at $s = 35$, the derived age ranges from $\sim 5 \times 10^8$ yrs using MCF90 (who adjust also the parameters for the foreground extinction) up to $\sim 15 \times 10^8$ yrs using CBB88. Based on the adopted calibrations, we show in Fig. 13 (a) and (b) the values of $L_{Bol}^{RGB}/L_T$ against age versus the "classical" (panel $a$) and "overshooting" (panel $b$) models. For the two models, $s = 35$ corresponds to an age of $6.5 \times 10^8$ and $11.9 \times 10^8$ yrs, respectively. We estimated the timescale for the full development of the RGB as the interval

$$\Delta t_{Ph-t} = t_{s=37} - t_{s=35}, \quad (3)$$

(see Figs. 11-13). For the "classical" and "overshooting" models, $t(s = 37)$ is $9.4 \times 10^8$ and $16.2 \times 10^8$ yr, respectively, yielding $\Delta t_{Ph-t} = 2.9 \times 10^8$ and $4.3 \times 10^8$. This result shows that passing from "classical" to "overshooting" models increases both the age at which the RGB ph-t takes place and the duration of the RGB ph-t . Such a long duration of the RGB ph-t means that use of the RGB ph-t for cosmological purposes would be problematic. Further effects in complex populations like galaxies, which in addition require cosmological corrections (*e.g.* $k-$ and $e-$corrections, see BCF93), means that the RGB ph-t probably could not be detected even in a galaxy with a single-burst population.

**5.3.2.** *Comparison with evolutionary models*

In the following discussion, we will compare our observations to two sets of theoretical models. The first set, was computed by SGR89,90 and is based on the classical treatment of mixing. The second set of models, presented by Al93, and BCF93, includes both classical models and those with a "mild overshooting".

Various other sets of similar computations have been carried out in this mass range (Castellani et al. 1990,1992; Maeder and Meynet 1991; Lattanzio 1991; Schaller et al. 1992, Schaerer et al. 1992) and, though different in many respects, their use would not alter the essence of our conclusions. In particular, the results concerning the difference in time-scale between "classical" and "overshooting" models, and the predictions concerning the behaviours of the fractional light contributions remain essentially the same.

Before describing the procedure we adopted to compare observations and models, we have however to analyse briefly how one could define the "observables" using the theoretical tracks as confusion on this item could give rise to misleading conclusions.

From a theoretical point of view, the epoch at which the RGB ph-t takes place must be defined clearly. In fact, while in old clusters the core mass and luminosity at the RGB-tip keep approximately constant as the age increases, with going younger across the transition they both decrease fairly rapidly, reaching a minimum, and then increase. This non-monotonic behaviour has been studied in detail by SGR89,90 through the computation of a fine grid of evolutionary sequences with canonical input physics, for different chemical compositions. These authors find that the variation in the RGB tip core mass and luminosity sets in at turn-off ages of $\sim 8 \times 10^8$ yr, while the RGB-tip minimum is reached at $\sim 4 \times 10^8$ yr, almost independent of composition.

Therefore, depending on which definition one adopts for the description of the RGB ph-t , one could obtain a different typical age for the RGB ph-t . The simplest definition is to adopt the mean value, $6 \pm 2 \times 10^8$ yr.

The models computed by Al93 without considering overshooting predict essentially the same evolutionary lifetimes for clusters undergoing the RGB ph-t : the variation in the RGB-tip luminosity occurs at cluster ages between $\sim 9$ and $\sim 4 \times 10^8$ yr.

Using the same models, but with overshooting, the RGB ph-t occurs at substan-tially older ages: the tip luminosity drops for ages in the range $\sim 1.9 - 1.3 \times 10^8$ yr. These figurs are slightly larger than the estimates we got from the $s-$age calibration, but there are two reasons able to explain the difference.

First, it is important to stress that the minimum luminosity of the RGB-tip in the SGR89,90 computations (in good agreement with the Al93 models with overshooting) is at Log L/L$_\odot$ $\sim$ 2.3 (for a mass of 2.5M$\odot$ and 1.8M$\odot$ for SGR89,90 and Al93 respectively). This luminosity is well below the limiting magnitude of the observed sample adopted here, $K_0 < 14.3$, which corresponds to Log L/L$_\odot$ $> 2.66$ (assuming $BC_K = 2.36$ at $(J-K)_0 = 0.75$). According to the models, the RGB brightens above Log L/L$_\odot$ $= 2.66$ only for cluster older than $\sim 5 \times 10^8$ (SGR89,90), $\sim 7 \times 10^8$ (Al93, no overshooting), $\sim 15 \times 10^8$ (Al93, with overshooting). As a consequence, with our observations we **cannot** test completely the whole theoretical extension of the RGB ph-t , but only its *old* portion, *i.e.* the part when the RGB stars begin to contribute significantly to the integrated cluster light.

Second, the calibration reported at Eq. 2 has been obtained by Chiosi et al. 1993 by estimating the age of a small set of calibrating clusters and it is therefore uncertain. Moreover, there is probably not a perfect correspondence between the specific model we have considered (with Z=0.008 and Y=0.25) and the grid of clusters they used.

Two other questions may be answered with our data: (i) What is the fractional contribution $L_{Bol}^{RGB}/L_T$ actually predicted by adopted models? (ii) How long is the "duration" of the RGB ph-t as obtained from the models compared with that determined from our data in Sect. 5.4.1.

To examine the first item, we have the theoretical models to compute the quantity $L_{Bol}^{RGB}/L_T$ directly from the tracks based on the framework of the Fuel Consumption Theorem discussed by RB86. Schematically from RB86, the fractional contribution can be computed by inserting the appropriate values in the formula:

$$L_{Bol}^{RGB}/L_T = 9.75 \times 10^{10} B(t) F^{RGB}, \tag{4}$$

where $F^{RGB}$ is the fuel burned during the considered evolutionary path. This quantity can be obtained directly from the tables in SGR89, while for Al93 models we compute for the various masses M$_i$

$$F^{RGB} = (Q^1_{H,tip} - Q^1_{H,2.66}) \times M_i \times X, \tag{5}$$

where X is the Hydrogen abundance and $Q^1_{H,tip}$ and $Q^1_{H,2.66}$ are the fractional masses of the inner border of the H-rich region at the RGB-tip and at Log $L/L_\odot = 2.66$ (our luminosity limit, see above), respectively (see also the legend of the Tables in Al93). The appropriate values for $B(t)$, which varies from $1.4 \times 10^{-11}$ up to $2.0 \times 10^{-11}$ over the interval $8.25 < Log\ t < 9.55$ have been taken from Guastamacchia (1992). Table 6 reports all the useful quantities involved in these calculations along with references to published material. The resulting figures for $L^{RGB}_{Bol}/L_T$ are then superimposed to the data in Fig. $13a, b$. Note that the time-scale used to plot the model predictions is the one directly read on the models, while the observational data (referring to each individual cluster) are based on the corresponding adopted $s-$age calibration. From the inspection of the diagrams shown in Fig. $13a, b$ we conclude:

a) The fractional light contributions predicted by the models (*i.e.* the values ranging from $\sim 0\%$ up to $\sim 17\%$) are in agreement with the observations if one takes into account the possible range spanned by $B(t)$.

b) Both the absolute values and the overall morphology of the RGB ph-t in the plane considered in Fig. 13 are independent of the treatment of mixing. In other words, as far as this specific aspect is concerned, there is no difference with passing from "classical" to "overshooting" models.

c) The slight discrepancy on the age axis is not reflecting a discrepancy between models and observations, but results from an imperfect match of the timescales directly obtained from the models and those obtained via the $s-$age calibration of a small set of MC clusters.

## 5.4 Playing with individual stars and integrated colours

### 5.4.1 *A preliminary test*

Since we have repeatedly noted that most of the integrated cluster light is actually contributed by the bright red stars we have observed, it is worthwhile carrying out a test to compare the integrated K-magnitudes available from the literature and those we can compute here by simply adding together the contributions of all the stars we measured in each cluster. We have therefore computed, for each cluster, the integrated K-magnitude and (J-K) colour obtained by adding the individual stars we plotted in Fig. 4 and then compared these values with those adopted from the literature (see Table 1). The results of this test are reported in Table 7 and the residuals in magnitude and colour (*literature − our value*) are plotted as a function of the $s$-parameter in Fig. $14a, b$, respectively.

The analysis of this figure can be very instructive and it gives moreover a hint on the size of the possible errors and statistical fluctuations.

In panel $(a)$, most clusters are within the range $\pm 0.3$ while few clusters show large residuals, and the overall distribution gives an apparent sloping trend with varying $s$ which is indeed partially spurious. For example NGC 2173 is $\sim 0.7$ magnitude fainter than our present extimate. This actually is due to the fact that a $\sim 30''$ diaphragm was used while our observations cover, in general a region of $\sim 60''$ radius. On the other hand, since our observations are not always precisely pointing the cluster centre (hard to define before observations for instance in NGC 2108, NGC 1831, NGC 2107), in low luminosity clusters with few bright red stars our integrated magnitudes may refer to quite different areas of the sky and the inclusion or not of only one bright (field) star affect significantly the integrated magnitude. This fact explains why our magnitudes for the clusters having just a few bright giants are smaller than the corresponding figures obtained by using large diaphragms.

The agreement in the integrated colours (Fig. $14b$) is better, even though a similar trend with varying $s$ is visible. However, the explanation of the differences is simple as we know that in computing our integrated K-magnitudes we have taken into account exclusively the red giants while, in the younger clusters ($s < 35$) the TO region is particularly bright and blue, contributing significantly to the integrated colours (see for example the case of NGC 1831).

#### **5.4.2** *The impact of AGB and RGB stars an integrated colours*

The last experiment is to understand (using the available sample of stars measured in BV in *Paper I* and in JHK here) how the integrated colours vary with varying the contributions of AGB and RGB stars in order to quantify the impact on the clusters integrated colours due to both AGB ph-t and RGB ph-t .

We have first adopted the integrated magnitudes (see below) of the various clusters and then have computed the new integrated magnitude of each cluster in each band after taking out the contribution of the AGB and RGB stars. Unfortunately, this simulation cannot be carried out homogeneously for all clusters in our sample because of the lack of full multicolour (B,V,J,H,K) photometry for all the cluster stars. We have therefore applied the procedure using different cluster subsets to study the impact of AGB and RGB stars on different colours:

(J-K) In the previous section we have shown that $(J-K)_{our}$ colour obtained adding the contribution of all the stars detected in our survey is compatible with the integrated J-K colours listed in Table 1. So that we choose (for consistency) to study the impact of AGB and RGB stars starting from $(J-K)_{our}$. The integrated J-K colour assumed and the colours obtained after the deletion of the AGB and RGB stars have been reported in column 2,3 and 4 of Table 8, respectively.

(V-K) Only five clusters in our sample (namely NGC 1831, NGC 2108, NGC 2162, NGC 2173, NGC 2209) have K and V magnitudes for most AGB and RGB stars. Only these clusters have been used to study the contribution of AGB and RGB stars to the V-K colour. For them, we have adopted the integrated V-K colours of the various clusters listed in Table 1. The integrated V-K colour assumed and the colours obtained after the deletion of the AGB and RGB stars have been reported in column 5,6 and 7 of Table 8, respectively.

(B-V) The impact of both red branches, $AGB + RGB$, on this colour has been studied using the B,V photometry presented in *Paper I.* By inspecting the optical CMDs, stars with $15 < V < 18.5$ and $B - V > 1.0$ have been assumed to be AGB or RGB stars while stars fainter or bluer are actually considered HB-clump and MS members. The integrated B-V colours, before and after the deletion of the red stars, have been reported in column 8 and 9 of Table 8, respectively.

To summarize the results, Fig. 15 shows the various colours before (full dots) and after the subtraction of AGB stars (open triangles) and of AGB + RGB stars (open squares). From Fig. 15 we conclude that:

a) The actual impact of the bright AGB stars on J-K cannot be constrained from the present sample as the inclusion of just one AGB star (sometimes almost as bright as the whole residual cluster) strongly modifies the result. For instance in NGC 2209, three bright stars (two very red and one intermediate) actually control the integrated cluster colour. Similarly, in V-K the importance of AGB stars is so strong that the exclusion of a few AGB objects decreases systematically the colours and also flattens the overall trend. In particular, the [V-K vs $s$]-plot yields a quite convincing confirmation that the V-K colour is essentially driven by the few bright AGB stars.

b) Since they are almost totally lacking in the CMDs, the exclusion of RGB stars is irrelevant for clusters with $s < 35$. With increasing $s$, the impact of this exclusion increases. However, though they are quite numerous especially in the clusters with $s > 38$, the presence or absence of RGB stars is not the dominant factor either in V-K or in B-V. In other words, they are too faint and with too intermediate colours to be able to dominate at any level the integrated colours. This implies that, although it is possible to detect the RGB ph-t in the fractional contribution to the K-integrated light (see Sect. 5.4), it is more difficult to detect its effect on the cluster integrated colours. Looking at the data of NGC 2162 and 2173 in Fig. 15*b*, one sees that the variation in integrated V-K caused by the RGB stars is noticeable, but it is nevertheless smaller than that due to the AGB.

c) Concerning specifically the behaviour of the integrated B-V colours, it seems evident from the plot in Fig. 15*c* that with increasing $s$, the population of the red stars increases and, correspondingly, the impact of their exclusion on the integrated B-V colour increases. The trend is still visible after the removal of *all* the red stars located in the above delimited area, even though it is affected by a strong scatter (see in particular NGC 1987 and 2162). With increasing further $s$, the influence of these stars on the B-V integrated colour decreases, essentially because (see NGC 2173) the contribution of the tip of the MS is now so low (since age is now high, and the TO is faint and quite red) that the integrated B-V colour remains red. In this picture the discussed variation of about 0.4 mag observed passing from clusters with $s \sim 30$ to those with $s \sim 40$ is *not* due to bright AGB and RGB stars, or, at least, it is so in a very partial way. For instance, in NGC 2173 the exclusion of these bright red components leads to a reduction in the B-V colour of a few hundredths of a magnitude. This implies that there is at least another component not considered here yet which partially controls the B-V colours.

From the above considerations and from the results shown in Fig. 15, we can thus conclude that the variations of the integrated colours of the MC clusters with $s = 31 - 43$ are controlled by the complex interplay of various factors, different from colour to colour and frequently dominated by the stochastic noise induced by few very bright objects. The overall picture emerging is consistent with the early conclusions drawn by PACFM83 and FMB90 that the J-K colour is mostly driven by the AGB stars, V-K is substantially controlled by AGB amd RGB (AGB stars being slightly more important), and, finally, B-V is partially influenced by the whole population of red stars brighter than the bulk of the RGB-clump but is also quite strongly dependent on the progressive fading and reddening of the turnoff stars (due to age increase).

## 6. SUMMARY AND CONCLUSIONS

In the present paper and in the companion *Paper I* (dealing with BV CCD-data) we report the results of the first step of our long-term project devoted to the detailed observational analysis of the stellar populations in a sample of MC clusters to test the stellar evolutionary models and to study the evolution of the integrated magnitudes and colours of *template simple stellar populations* (SSP) for cosmological purposes.

In particular, using the available BVJHK data (11 LMC clusters observed in BV, 12 in JHK, 9 in common, about 20000 stars measured in total), we have studied two specific problems, also carrying out detailed comparisons with the theoretical model predictions and useful tests and simulations.

The first item we have dealt with is the study of the existence and complete description via observational quantities of the so-called AGB ph-t and RGB ph-t (RB86).

The second problem is the analysis of the possible impact of these ph-ts , and especially of the bright AGB and RGB stars, on integrated SSP magnitudes and colours.

The near-IR observations, carried out with a $64 \times 58$ array at CTIO, covered the central regions of the clusters with 4-frame mosaics and posed many difficult reduction problems due to crowding and undersampling. In this respect, the use of new larger and better devices would simplify any further study, and is to be recommended.

Though problems like small number of observed clusters, statistical fluctuations, background contamination, difficulties in assembling complete BVJHK-catalogues, etc. require further data, there are in our view some direct observational indications which (i) give support to the overall theoretical framework, (ii) essentially confirm the existence of the predicted *phase transitions* (RB86), and (iii) allow us to evaluate their impact on the integrated magnitudes and colours, at least to a first approximation.

More specifically, the main results of the present work are:

1. The presentation of near-IR CMD's down to K=16 for 12 MC clusters, with a total sample of $\sim$ 450 stars.

2. The overall confirmation of the results presented from the early near-IR studies by PACFM83 and FMB90 concerning the highlights of the interpretation of the integrated magnitudes and colours of the MC clusters.
3. The direct observational verification of the existence of the so-called RGB ph-t predicted by RB86, using both star counts and fractional light contributions. In fact, though in general dominated by the statistical fluctuations associated with the prevailing (in brightness) AGB stars, it has been feasible to demonstrate the existence and effects of the predicted RGB ph-t , after removing the AGB stars detected in each cluster sample.
4. Concerning the AGB and bright RGB (Log $L/L_{\odot} \sim 2.66$) overall contribution to the cluster K-integrated light, they yield about 70% of the total or even more. The AGB contribution is in these clusters highly fluctuating, but always close to $\sim$ 60%. The RGB contribution varies with $s$ (in turn, age) as predicted by the models, ranging from $\sim$ 0% up to $\sim$ 17% for $s > 35$.
5. The age at which the RGB ph-t actually takes place depends on the adopted $s-$ vs age calibration, and, in turn, on the adopted theoretical models. Using "classical models", the RGB ph-t occurs at $\sim 6 \pm 2 \times 10^8$ yr, where the associated uncertainty reflects, on the one hand, the difficulty to determine precisely the ph-ts ages (both in the observational and the theoretical planes) and, on the other, the plausible duration of the transition itself, *i.e.* the time necessary to go in the CMD of a given cluster from the appearance of the first RGB stars up to the full development of the whole RGB. Passing to "overshooting" models, the essence of the phenomenon is the same, but there is a time-shift to $\sim 15 \pm 3 \times 10^8$ yr, with a corresponding increase of the duration. Note that a longer duration of the RGB ph-t makes it less useful as a possible time-mark for cosmological purposes, all the other effects being kept constant.
6. The quantitative comparisons with the models show further that, while the age and duration of the RGB ph-t depends on the treatment of mixing, "classical" and "overshooting" models yields exactly the same figures for the fractional contributions of RGB stars to the total integrated cluster light. This can be easily explained based on the "Fuel Cunsumption Theorem" (RB86), as the total fuel actually burned during the RGB-phase is the same in the two cases.

**Acknowledgements**

This work is based on observations obtained at the Cerro Tololo Inter-American Observatory.

Cerro Tololo Inter-American Observatory, National Optical Astronomy Observatories, are operated by the Association of Universities for Research in Astronomy, Inc. (AURA) under cooperative agreement with the National Science Foundation.

This research was supported by the *Consiglio delle Ricerche Astronomiche* (CRA) and the *Comitato Universitario Nazionale* (CUN) of the *Ministero dell' Universitá e della Ricerca Scientifica e Tecnologica* (MURST). We thank the CTIO staff for their assistance in acquiring the IR-data discussed here, and in particular Dr. Jay Elias. Special thanks are due to Alvio Renzini for inspring the whole project , to Cesare Chiosi and Alessandro Bressan for helpful discussions, and to Livia Origlia for assistance during the data reduction.

**Figure Captions**

**Fig. 1 a), b), c), d)** Integrated colours *vs.* s-parameter. Large full dots indicate the clusters in our sample, which are labelled with their NGC numbers.

**Fig. 2 a), b), c)** Calibrating relations in J, H, K respectively.

**Fig. 3 a) – l)** Computer maps for the clusters in our sample. Coordinates are in pixels.

**Figure 4 a), b), c)** Plot of the residuals between our data and the literature for all the identified stars. The differences are in the sense: literature – this work. Different symbols are for different clusters: open squares: NGC 1783, filled circles: NGC 1978, filled squares: NGC 1806, open circles: NGC 1831, five-pointed stars: NGC 1987, open triangles: NGC 2162, filled triangles: NGC 2173, crosses: NGC 2209, asterisks: NGC 2108, eight-pointed stars: NGC 1756.

**Fig. 5** C–M diagrams $(K, J-K)$ for the twelve clusters shown in order of increasing NGC number.

**Fig. 6 a), b)** Cumulative C–M diagrams for the 12 clusters, divided in two sets (see text).

**Fig. 7** Colour-colour diagram for all the observed stars brighter than $K = 14.3$. The loci drawn for Carbon stars and LPVs (long-dashed) are from Bessell and Brett, 1988, and the short dashed for LPVs with $P > 350^d$ is from FPAM78. The mean loci for K giants by FPAM78 are also plotted (solid line).

**Fig. 8 a), b)** The same as Fig. 7 dividing the clusters as in Fig. 6.

**Fig. 9** Cumulative $(K_0, (V-K)_0)$diagram for the twelve clusters. The mean ridge line is drawn.

**Fig. 10 a), b), c)** Plot of $N_4$ *vs* s. Panel a) is for all the AGB and bright RGB stars in each cluster, panel b) is for the AGB only, panel c) for the bright RGB only. The numbers are normalised to the total luminosity of the clusters in units of $10^4 L_\odot$(see text).

**Fig. 11 a), b)** Contributions to the total K magnitude of each cluster. Panel a) is for AGB, panels b) is for the bright RGB.

**Fig. 12** The same as Fig. 11 c), but for the bolometric magnitude. The dashed lines represent the average contribution from NGC 1831, 2107, 1756, 1868, 2209 ($L_{bol}^{RGB}/L_T \sim 0.007$) on one hand, and NGC 1783, 1806, 1978, 2173, 2162 ($L_{bol}^{RGB}/L_T \sim 0.176$) on the other hand.

**Fig. 13 a), b)** Comparisons with theoretical models. Panel a) - classical models; the $s-$age calibration is from EF88. The dashed line represents the theoretical expectations (see Sect. 5.3). Panel b) - overshooting models; the $s-$age relation is from Chiosi et al. (1993) (see Sect. 5.3.1. Eq. 2a).

**Fig. 14 a), b)** Differences in magnitude and colour between the values listed in Table 1 and the sum of the contribution of the resolved stars measured in this work. Empty triangles in panel a) indicates the clusters having magnitudes obtained with a $30''$ diaphragm aperture.

**Fig. 15 a), b), c)** Contribution of each evolutionary phase to the cluster integrated colour. Dots indicate the colour reported in literature, triangles the residual colour after AGB star removal, squares the colour after bright AGB + RGB star removal.

## REFERENCES

Aaronson M., Mould J.R., 1982, ApJS, 48, 161 (AMMAII).

Alongi M., Bertelli G., Bressan A., Chiosi C., Fagotto F., Greggio L., Nasi E., 1993, A&AS, 97, 851. (Al93)

Arimoto N., Bica E., 1989, A&A, 222, 89.

Baade W., 1951, Publ. Univ. Michigan Observ., 10, 7.

Barbaro G., Olivi F.M., 1991, AJ, 101, 922.

Barbero J., Brocato E., Cassatella A., Castellani V., Geyer E.H., 1990, ApJ, 351, 98.

Battinelli P., Capuzzo-Dolcetta R., 1989, ApJ, 347, 794.

Bessell M.S., Brett J.M., 1988, PASP, 100, 1134.

Bica E., Dottori H., Pastoriza M., 1986, A&A, 156, 261. (BDP86)

Bica E., Alloin D., Santos J.F.C., 1990, A&A, 235, 103.

Bica E., Claria J.J., Dottori H., Santos J., Piatti A., 1991, ApJ, 381, L51. (BCDSP91)

Bica E., Claria J.J., Dottori H., 1992, AJ, 103, 1859.

Bressan A., Chiosi C., Fagotto F., 1993, preprint. (BCF93)

Brocato E., Buonanno R., Castellani V., Walker A.R., 1989, ApJS, 71, 25.

Brunet J.P., 1975, A&A, 43, 345.

Bruzual G.A., Charlot S., 1993, ApJ, 405, 538.

Buonanno R., Corsi C.E., De Biase G.A., Ferraro I., 1979, in Sedmak G., Capaccioli M., Allen R.J., eds., Image Processing in Astronomy, Trieste Observatory, p.354.

Buonanno R., Buscema G., Corsi C.E., Ferraro I., Iannicola G., 1983, A&A, 126, 278.

Buonanno R., Iannicola G., 1989, PASP, 101, 294.

Burstein D., Heiles C., 1982, AJ, 87, 1165.

Castellani V., Chieffi A., Straniero O., 1990, ApJS, 74, 463.

Castellani V., Chieffi A., Straniero O., 1992, ApJS, 78, 517.

Chambers K.C., Charlot S. 1990, ApJ, 348, L1.

Charlot S., Bruzual G., 1991, ApJ, 367, 126.

Chiosi C., Pigatto L., 1986, ApJ, 308, 1.

Chiosi C., Bertelli G., Bressan A., Nasi E., 1986, A&A, 165, 84.

Chiosi C., Bertelli G., Bressan A., 1988, A&A, 196,84. (CBB88)

Chiosi C., Bertelli G., Bressan A., 1993, private communication.

Chokshi A., Wright E.L., 1987, ApJ, 319, 44.

Cohen J.G., 1982, ApJ, 258, 143.

Cohen J.G., Frogel J.A., Persson S.E., Elias J.H., 1981, ApJ, 249, 481. (CFPE81)

Corsi C.E., Testa V., 1992, in Brocato E., Ferraro F.R., Straniero O., eds., Star Clusters and Stellar Evolution, Mem. S. A. It., vol. 63, n.1, p. 131.

Corsi C.E., Buonanno R., Ferraro F.R., Fusi Pecci F., Greggio L., Testa V., 1994, MNRAS, in press. (*Paper I* )

Danziger I.J., 1973, ApJ, 181, 641.

Davidge T.J., Harris W.E., Bridges T.J., Hanes D.A., 1992, AJ, 81, 251.

Dottori H., Melnick J., Bica E., 1987, Rev. Mex. Astron. Astrof., 14, 183.

Elias J.H., Frogel J.A., Matthews K., Neugebauer G., 1982, AJ, 87, 1029.

Elson R.A.W., Fall S.M., 1985, ApJ, 299, 211. (EF85)

Elson R.A.W., Fall S.M., 1988, AJ, 96, 1383. (EF88)

Ferraro F.R., Clementini G., Fusi Pecci F., Buonanno R., 1990, A&AS, 84, 59.

Ferraro F.R., Fusi Pecci F., Testa V., 1994, in Ian S. McLean ed., Infrared Astronomy with Arrays: the Next Generation, Kluwer, Dordrecht, p. 127.

Fischer P., Welch D.L., Mateo M., 1992, AJ, 104, 1086.

Freeman K.C., Illingworth G., Oemler A., 1983, AJ, 272, 488.

Frogel J.A., 1984, PASP, 96, 856.

Frogel J.A., 1988, ARAA, 26, 51.

Frogel J.A., Persson S.E., Aaronson M., Matthews K., 1978, ApJ, 220, 75. (FPAM78)

Frogel J.A., Persson S.E., Cohen J.G., 1981, ApJ, 246, 842. (FPC81)

Frogel J.A., Cohen J.G., 1982, ApJ, 253, 580.

Frogel J.A., Cohen J.G., Persson S.E., 1983, ApJ, 275, 773. (FCP83)

Frogel J.A., Elias J.H., 1988, ApJ, 324, 823. (FE88)

Frogel J.A., Mould J.R., Blanco V.M. 1990, ApJ, 352, 96. (FMB90)

Fusi Pecci F., Cacciari C., 1991, in Sanchez F., Vasquez M., eds., New Windows to the Universe, Cambridge: Cambridge Univ. Press, p.335.

Gascoigne S.C.B., 1962, MNRAS, 124, 201.

Gascoigne S.C.B., 1971, in Muller A.B., ed., The Magellanic Clouds, Reidel, Dordrecht, p. 25.

Gascoigne S.C.B., 1980, in J.E. Hesser, ed., IAU Symp. No. 85, Star Clusters, Reidel, Dordrecht, p. 305.

Gascoigne S.C.B., Kron G.E., 1952 PASP, 64, 196.

Geyer E.H., Hopp U., 1982, Ap&SS, 84, 133.

Girardi L., Bica E., 1992, A&A, 274, 279.

Greggio L., 1987, in Azzopardi M., Matteucci F., eds., Stellar Evolution and Dynamics in the Outer Halo of the Galaxy, Garching: ESO, p. 453.

Guastamacchia M., 1992, Master Thesis, Univ. of Bologna.

Hodge P.W., 1983, ApJ, 264, 470.

Iben I.Jr, Renzini A., 1983, ARAA, 21, 271.

Kontizas M., Chrysovergis M., Kontizas E., 1987, A&AS, 68, 147.

Lattanzio J.C., 1991, ApJS, 76, 215.

Lloyd-Evans T.L.E., 1980, MNRAS, 193, 87. (LE80)

Maeder A., Meynet G., 1991, A&AS, 89, 451.

Mateo M., 1987, ApJ, 323, L41.

Mateo M., 1989, Ap&SS, 156, 85.

Mateo M., 1992, in Barbuy B., Renzini A., eds., The Stellar Populations of Galaxies, IAU Symp. No. 149, Kluwer, Dordrecht, p. 147.

McCaughrean M., 1989, in Third AMES Workshop on Infrared Detector Technology.

Meurer G.R., Cacciari C., Freeman K.C., 1990, AJ, 99, 1124 (MCF90).

Mould J., Aaronson M., 1979, ApJ, 232, 421. (MA79)

Mould J.R., Aaronson M., 1980, ApJ, 240, 464 (AMMAI).

Mould J.R., Aaronson M., 1982, ApJ, 263, 629 (AMMAIII).

Mould J.R., Da Costa G.S., Wieland F.P., 1986, ApJ, 309, 39.

Mould J.R., Da Costa G.S., 1988, in Blanco V.M., Philips M.M., eds., Progress and Opportunities in Southern Hemisphere Astronomy, Brigham Young University, p. 197.

Mould J.R., Kristian J., Nemec J., Aaronson M., Jensen J., 1989, ApJ, 339, 84.

Olszewski E.W., 1984, ApJ, 284,108.

Olszewski E.W., Schommer R.A., Suntzeff N.B., Harris H.C., 1991, AJ, 101, 515. (OSSH91)

Öpik E., 1938, Publ. Obs. Tartu, 30, No. 3,4.

Persson S.E., Aaronson M., Cohen J.G., Frogel J.A., Matthews K., 1983, ApJ, 266, 105. (PACFM83)

Renzini A., 1981, Ann. Phys. Fr. 6, 87.

Renzini A., 1991, in Shanks T., Banday A.J., Ellis R.S., Frenk C.S. and Wolfendale A.W., eds., 1991, Observational Tests of Cosmological Inflation, NATO ASI Series

C, vol. 348, p. 131.

Renzini A. 1992, in B. Barbuy, A. Renzini, eds., IAU Symp. No. 149, The Stellar Population of Galaxies, Kluwer, Dordrecht, p. 325.

Renzini A., Buzzoni A., 1983, Mem. Soc. Astron. It., 54 739.

Renzini A., Buzzoni A., 1986, in C. Chiosi, A. Renzini, eds., Spectral Evolution of Galaxies, Reidel, Dordrecht, p. 195. (RB86)

Sandage A.R., Eggen O.J., 1960, MNRAS, 121, 232.

Sandage A. R., Schwarzschild A., 1952, ApJ, 116, 463.

Savage B.D., Mathis J.S., 1979, ARAA, 17, 73.

Schaerer D., Meynet G., Maeder A., Schaller G., 1993, A&AS, 98, 523.

Schaller G., Schaerer D., Meynet G., Maeder A., 1992, in press.

Schommer R.A, Olszewski E.W., Aaronson M., 1984, ApJ, 285, L53.

Searle L., Wilkinson A., Bagnuolo W.G., 1980, ApJ, 239, 803. (SWB)

Seggewiss W., Richtler T., 1989, in de Boer K.S., Spite F., Stasinka G., eds., Recent Developments of MC Research, Obs. de Paris, p. 45.

Sweigart A.V., Greggio L., Renzini A., 1989, ApJS, 69, 911.

Sweigart A.V., Greggio L., Renzini A., 1990, ApJ, 364, 527.

Vallenari A., Chiosi C., Bertelli G., Meylan G., Ortolani S., 1992, AJ, 104, 1100.

van den Bergh S., 1981, A&AS, 46, 79. (vdB81)

Westerlund B.E., 1990, A&ARev, 2, 29.

Wyse R.F.G., 1985, ApJ, 299, 593.

**Table 1.** IR Photometric data from literature.

| Cluster | $K_{int}$ | $(V-K)_0$ | $(J-K)_0$ | $(H-K)_0$ | $(B-V)_0$ | $E(B-V)$ | $SWB$ | $s$ |
|---|---|---|---|---|---|---|---|---|
| NGC 1756 | | | | | 0.40[16] | | | 32[2] |
| NGC 1783 | 9.12(30)[3] | 2.33[3] | 0.74[3] | 0.13[3] | 0.62[16] | 0.10[5] | *V*[1,4] | 37[2] |
| | 8.65(56)[3] | | | | | 0.06[13] | | 38.0[5] |
| | 8.56(60)[3] | | | | | | | |
| NGC 1806 | 9.09(24)[3] | 2.79[3] | 0.89[3] | 0.22[3] | 0.73[16] | 0.12[3] | *V*[1] | 40[2] |
| | 9.02(29)[3] | | | | | | | |
| | 8.44(56)[3] | | | | | | | |
| | 8.19(60)[3] | | | | | | | |
| NGC 1831 | 10.71(30)[3] | 1.40[3] | 0.48[3] | 0.18[3] | 0.34[16] | 0.10[3] | *V*[1] | 31[2] |
| | 9.21(59)[3] | | | | 0.35[17] | 0.05[15] | | 32.7[5] |
| | | | | | | 0.04[13] | | |
| | | | | | | 0.07[5] | | |
| NGC 1868 | 9.97(24)[3] | 1.63[3] | 0.69[3] | 0.15[3] | 0.45[16] | 0.07[3] | *IV*[4,7] | 33[2] |
| | 9.73(64)[3] | | | | | | | 34.5[5] |
| NGC 1978 | 8.44(30)[3] | 2.58[3] | 0.93[3] | 0.27[3] | 0.78[16] | 0.10[3] | *VI*[1] | 45[2] |
| | 7.86(56)[3] | | | | | 0.07[5] | | |
| | 7.92(60)[3] | | | | | 0.19[11] | | |
| NGC 1987 | 10.00(24)[3] | 2.69[3] | 0.89[3] | 0.22[3] | 0.52[16] | 0.12[3,6] | *IV*[1] | 35[2] |
| | 9.87(30)[3] | | | | | | | 35.1[5] |
| | 9.01(60)[3] | | | | | | | |
| NGC 2107 | 9.53(30)[3] | 1.93[3] | 0.80[3] | 0.20[3] | 0.38[16] | 0.19[3] | *IV*[1] | 32[2] |
| | 9.21(60)[3] | | | | | | | |
| NGC 2108 | 9.66(24)[3] | 2.54[3] | 1.18[3] | 0.48[3] | 0.58[16] | 0.18[3] | *IV − V*[7] | 36[2] |
| | 9.27(64)[3] | | | | | | | |
| NGC 2162 | 10.39(30)[3] | | 0.95[3] | 0.20[3] | 0.68[16] | 0.07[3] | *V*[1] | 39[2] |
| | | | | | | 0.05[5] | | 40.5[5] |
| | | | | | | 0.04[8] | | |
| | | | | | | 0.06[9] | | |
| NGC 2173 | 9.90(30)[3] | 2.90[3] | 1.04[3] | 0.24[3] | 0.84[16] | 0.07[3,11] | *V − VI*[1] | 42[2] |
| | | | | | 0.86[16] | 0.12[10] | *VI*[4,7] | 42.5[5] |
| NGC 2190 | | | | | | 0.10[8] | | |
| NGC 2209 | 10.04(30)[3] | | 1.68[3] | 0.66[3] | 0.53[16] | 0.07[3] | *III − IV*[1] | 35[2] |
| | | | | | | 0.15[14] | *IV*[7] | 36.9[5] |
| | | | | | | 0.06[5] | | |
| NGC 2249 | | | | | 0.43[7] | 0.12[5] | | 34[2] |
| | | | | | 0.39[18] | 0.10[12] | | 33.6[5] |
| | | | | | 0.42[18] | | | |

References: see below
Number in parenthesis in K column are the sizes, in arcseconds, of diaphgrams used for integrated photometry.

References: (1)- SWB, 1980; (2)- Elson and Fall, 1985; (3)- PACFM, 1983; (4)- Freeman, Illingworth and Oemler, 1983; (5)- Meurer, Cacciari and Freeman, 1990; (6)- Aaronson and Mould, 1982 (AMMA II); (7)- Bica, Dottori and Pastoriza, 1986; (8)- Schommer, Olszewski and Aaronson, 1986; (9)- Chiosi and Pigatto, 1986; (10)- Mould, Da Costa and Wieland, 1986; (11)- Mould and Da Costa, 1988; (12)- Burstein and Heiles, 1982; (13)- Westerlund, 1990; (14)- Dottori *et al.*, 1987; (15)- Vallenari *et al.*, 1992, (16)- van den Bergh, 1981; (17)- Mateo, 1987; (18)- Bica *et al.*, 1991.

**Table 2.** Magnitudes, colours and positions for the program stars in each cluster.

| *Name* | $K$ | $J-K$ | $H-K$ | $V-K$ | $B-V$ | $X$ | $Y$ |
|---|---|---|---|---|---|---|---|
| **NGC 1756** | | | | | | | |
| 1 | 15.19 | 0.50 | -0.22 | 1.89 | 0.81 | 35.42 | 11.88 |
| 2 | 14.79 | 0.59 | 1.67 | 3.14 | 1.32 | 34.83 | 15.15 |
| 3 | 15.41 | -0.60 | -0.94 | 1.57 | 1.05 | 22.18 | 18.68 |
| 4 | 14.37 | 0.51 | -0.18 | 2.64 | 0.58 | 40.73 | 19.83 |
| 5 | 14.43 | 1.01 | 0.62 | 3.06 | 0.82 | 24.83 | 19.94 |
| 6 | 14.90 | 0.46 | -0.04 | 2.37 | 0.95 | 36.31 | 20.72 |
| 7 | 14.35 | 0.46 | -0.01 | 2.61 | 0.79 | 39.09 | 21.66 |
| 8 | 14.89 | 0.71 | 0.03 | 3.26 | 1.38 | 20.47 | 22.83 |
| 9 | 14.67 | 0.32 | 0.05 | 2.23 | 0.84 | 31.93 | 23.03 |
| 10 | 14.38 | 0.80 | 0.11 | 3.25 | 1.30 | 48.83 | 23.09 |
| 11 | 14.17 | 0.65 | 0.05 | 3.18 | 1.39 | 34.64 | 23.47 |
| 12 | 15.10 | 0.33 | -0.09 | 2.37 | 0.77 | 46.69 | 26.53 |
| 13 | 14.53 | 0.64 | 0.49 | 2.82 | 0.99 | 30.64 | 28.31 |
| 14 | 14.11 | 0.46 | 0.13 | 2.82 | 1.21 | 21.92 | 28.78 |
| 15 | 15.29 | -0.16 | -0.32 | 2.09 | 0.78 | 3.96 | 31.42 |
| 16 | 14.69 | 0.06 | 0.03 | 2.02 | 0.46 | 32.59 | 31.67 |
| 17 | 14.67 | 0.65 | 0.01 | 3.29 | 1.27 | 11.56 | 33.15 |
| 18 | 13.39 | 0.97 | 0.16 | 3.93 | 0.00 | 30.98 | 34.25 |
| 19 | 14.26 | 0.34 | 0.00 | 2.38 | 0.59 | 38.30 | 35.47 |
| 20 | 14.64 | 0.43 | 0.05 | 2.57 | 1.08 | 44.95 | 35.92 |
| 21 | 14.38 | 0.88 | 0.08 | 3.20 | 1.35 | 48.15 | 37.64 |
| 22 | 14.73 | 0.56 | -0.05 | 2.64 | 0.97 | 24.94 | 38.70 |
| 23 | 14.53 | 0.40 | 0.27 | 2.29 | 0.74 | 30.92 | 40.66 |
| 24 | 14.37 | 0.66 | 0.41 | 2.57 | 0.92 | 17.41 | 43.05 |
| 25 | 11.73 | 0.81 | 0.26 | 4.89 | 1.50 | 36.36 | 48.95 |
| 26 | 14.35 | 0.86 | 0.01 | 3.57 | 1.47 | 11.66 | 55.19 |
| **NGC 1783** | | | | | | | |
| 1 | 11.37 | 1.15 | 0.26 | 4.86 | 1.53 | 97.18 | 3.17 |
| 2 | 12.88 | 0.95 | 0.17 | — | — | 56.25 | 12.31 |
| 3 | 14.71 | 0.51 | -0.09 | — | — | 31.18 | 15.99 |
| 4 | 14.24 | 0.34 | 0.12 | 1.39 | 0.53 | 93.64 | 17.19 |
| 5 | 15.21 | 0.87 | 0.26 | — | — | 46.12 | 21.34 |
| 6 | 14.12 | 0.84 | 0.17 | — | — | 24.11 | 22.88 |
| 7 | 14.62 | 0.51 | 0.07 | — | — | 46.42 | 23.16 |
| 8 | 14.81 | 0.70 | 0.01 | — | — | 32.67 | 25.45 |
| 9 | 13.88 | 0.65 | 0.03 | — | — | 40.60 | 25.54 |
| 10 | 14.09 | 0.74 | 0.15 | — | — | 18.06 | 26.75 |
| 11 | 13.11 | 0.91 | 0.11 | — | — | 32.33 | 29.27 |
| 12 | 14.83 | 0.47 | 0.28 | — | — | 47.53 | 30.24 |
| 13 | 14.98 | 0.55 | 0.03 | — | — | 42.97 | 31.32 |
| 14 | 15.15 | 0.48 | -0.22 | — | — | 30.27 | 32.64 |
| 15 | 13.10 | 0.81 | 0.10 | — | — | 27.27 | 33.22 |
| 16 | 13.80 | 0.76 | 0.05 | — | — | 11.02 | 34.12 |
| 17 | 14.75 | 0.68 | -0.20 | — | — | 51.83 | 35.48 |
| 18 | 14.01 | 0.52 | -0.02 | — | — | 47.81 | 35.89 |
| 19 | 11.39 | 1.17 | 0.20 | 4.48 | 1.78 | 89.58 | 36.37 |
| 20 | 13.69 | 0.81 | 0.08 | — | — | 26.88 | 36.96 |
| 21 | 14.64 | 0.78 | 0.14 | — | — | 50.76 | 37.74 |
| 22 | 14.75 | 0.59 | 0.12 | — | — | 73.64 | 39.37 |

**Table 2.** *(continued)*

| *Name* | $K$ | $J-K$ | $H-K$ | $V-K$ | $B-V$ | $X$ | $Y$ |
|---|---|---|---|---|---|---|---|
| 23 | 15.18 | 0.35 | 0.08 | — | — | 63.71 | 39.64 |
| 24 | 13.55 | 0.86 | 0.19 | — | — | 51.31 | 40.03 |
| 25 | 14.83 | 0.80 | 0.31 | — | — | 60.25 | 40.44 |
| 26 | 14.07 | 0.16 | 0.15 | — | — | 57.73 | 40.51 |
| 27 | 15.05 | 0.50 | -0.12 | — | — | 72.50 | 42.02 |
| 28 | 13.80 | 1.55 | 0.86 | — | — | 44.28 | 42.34 |
| 29 | 13.18 | 0.64 | 0.03 | — | — | 46.46 | 42.86 |
| 30 | 14.91 | 0.44 | 0.20 | — | — | 23.86 | 43.24 |
| 31 | 15.43 | 0.20 | — | — | — | 34.00 | 44.62 |
| 32 | 13.43 | 0.96 | 0.25 | — | — | 53.68 | 45.16 |
| 33 | 15.20 | 0.47 | -0.01 | — | — | 45.49 | 45.24 |
| 34 | 14.49 | 0.38 | 0.10 | — | — | 50.24 | 45.68 |
| 35 | 11.38 | 1.03 | 0.30 | — | — | 1.55 | 46.58 |
| 36 | 15.49 | 0.20 | — | — | — | 31.38 | 46.77 |
| 37 | 11.43 | 1.07 | 0.22 | — | — | 59.25 | 46.86 |
| 38 | 14.05 | 0.87 | 0.07 | — | — | 63.57 | 48.20 |
| 39 | 14.40 | 0.72 | 0.15 | — | — | 75.77 | 48.40 |
| 40 | 14.55 | 0.83 | 0.46 | — | — | 55.01 | 49.51 |
| 41 | 12.70 | 1.06 | 0.28 | — | — | 62.98 | 50.37 |
| 42 | 14.69 | 0.77 | 0.24 | — | — | 1.89 | 51.73 |
| 43 | 14.06 | 0.80 | 0.15 | — | — | 42.19 | 51.92 |
| 44 | 13.88 | 0.78 | 0.15 | — | — | 19.20 | 52.07 |
| 45 | 14.07 | 0.83 | 0.33 | — | — | 49.59 | 53.31 |
| 46 | 13.71 | 0.90 | 0.19 | — | — | 8.67 | 54.73 |
| 47 | 13.56 | 0.69 | 0.06 | — | — | 48.88 | 54.87 |
| 48 | 13.80 | 0.81 | 0.16 | 2.64 | 1.39 | 85.86 | 54.91 |
| 49 | 14.33 | 0.62 | -0.01 | 2.67 | 1.04 | 91.61 | 56.15 |
| 50 | 14.21 | 0.68 | 0.06 | — | — | 31.98 | 58.06 |
| 51 | 13.83 | 0.63 | 0.12 | — | — | 72.83 | 58.13 |
| 52 | 11.04 | 1.12 | 0.22 | — | — | 45.66 | 58.36 |
| 53 | 11.65 | 1.08 | 0.19 | — | — | 64.31 | 59.44 |
| 54 | 14.23 | 0.77 | 0.11 | — | — | 87.74 | 61.08 |
| 55 | 14.82 | 0.35 | -0.05 | — | — | 79.06 | 61.86 |
| 56 | 15.08 | 0.43 | -0.04 | — | — | 35.60 | 64.53 |
| 57 | 14.28 | 0.79 | 0.18 | — | — | 51.66 | 64.54 |
| 58 | 14.44 | 0.76 | 0.14 | — | — | 74.07 | 64.69 |
| 59 | 14.82 | 0.73 | 0.12 | — | — | 83.17 | 65.85 |
| 60 | 12.13 | 0.99 | 0.20 | — | — | 52.78 | 67.50 |
| 61 | 14.64 | 0.58 | 0.05 | — | — | 98.33 | 69.10 |
| 62 | 13.16 | 0.80 | 0.16 | 3.38 | 1.36 | 77.17 | 75.16 |
| 63 | 13.47 | 0.69 | 0.08 | — | — | 66.28 | 75.98 |
| 64 | 14.29 | 0.70 | 0.07 | 2.80 | 1.14 | 97.71 | 78.15 |
| 65 | 13.86 | 0.72 | 0.15 | 3.31 | 1.32 | 24.53 | 80.84 |
| 66 | 13.52 | 0.76 | 0.11 | 2.76 | 1.32 | 51.00 | 82.34 |
| 67 | 14.72 | 0.13 | -0.32 | — | — | 63.86 | 85.89 |
| 68 | 15.13 | 0.53 | -0.01 | — | — | 50.35 | 86.31 |
| 69 | 13.17 | 0.98 | 0.28 | 3.04 | 1.30 | 16.97 | 86.42 |
| 70 | 14.15 | 1.78 | 1.35 | — | — | 62.07 | 86.48 |
| 71 | 10.61 | 1.89 | 0.73 | 5.61 | 2.36 | 5.67 | 87.51 |
| 72 | 14.88 | 0.55 | 0.07 | — | — | 83.94 | 97.07 |

**Table 2.** *(continued)*

| *Name* | *K* | *J* − *K* | *H* − *K* | *V* − *K* | *B* − *V* | *X* | *Y* |
|---|---|---|---|---|---|---|---|
| **NGC 1806** | | | | | | | |
| 1 | 10.43 | 1.91 | 0.70 | — | — | 72.42 | 12.47 |
| 2 | 11.23 | 1.08 | 0.19 | — | — | 45.04 | 14.33 |
| 3 | 13.62 | 0.87 | 0.20 | — | — | 19.56 | 14.90 |
| 4 | 13.74 | 0.84 | 0.16 | — | — | 22.36 | 15.13 |
| 5 | 14.18 | 0.81 | 0.30 | — | — | 67.36 | 15.26 |
| 6 | 14.14 | 1.04 | 0.40 | — | — | 42.78 | 16.25 |
| 7 | 12.92 | 0.79 | 0.14 | — | — | 75.32 | 17.15 |
| 8 | 13.49 | 0.77 | 0.23 | — | — | 72.69 | 17.83 |
| 9 | 12.76 | 0.86 | 0.15 | — | — | 77.37 | 18.35 |
| 10 | 15.15 | 0.19 | -0.03 | — | — | 80.67 | 19.25 |
| 11 | 14.03 | 0.80 | 0.24 | — | — | 67.11 | 19.32 |
| 12 | 14.79 | 0.52 | 0.40 | — | — | 71.00 | 20.86 |
| 13 | 15.15 | 0.29 | 0.21 | — | — | 68.16 | 28.47 |
| 14 | 14.41 | 0.59 | 0.17 | — | — | 72.81 | 28.49 |
| 15 | 13.90 | 0.68 | 0.12 | — | — | 79.28 | 29.11 |
| 16 | 14.52 | 0.83 | 0.39 | — | — | 61.55 | 30.34 |
| 17 | 15.54 | 0.80 | 0.07 | — | — | 85.54 | 32.56 |
| 18 | 15.83 | 0.74 | 0.97 | — | — | 83.00 | 32.65 |
| 19 | 11.69 | 0.95 | 0.19 | — | — | 74.35 | 33.20 |
| 20 | 14.70 | 0.50 | -0.10 | — | — | 35.33 | 35.28 |
| 21 | 13.32 | 0.69 | 0.16 | — | — | 92.66 | 35.70 |
| 22 | 13.83 | 0.81 | 0.10 | — | — | 44.10 | 36.46 |
| 23 | 13.96 | 0.55 | 0.09 | — | — | 69.60 | 36.96 |
| 24 | 14.28 | 0.50 | 0.15 | — | — | 86.43 | 37.31 |
| 25 | 13.12 | 0.83 | 0.14 | — | — | 66.40 | 37.58 |
| 26 | 12.59 | 0.92 | 0.21 | — | — | 42.38 | 38.03 |
| 27 | 13.64 | 0.81 | 0.20 | — | — | 88.28 | 38.28 |
| 28 | 15.15 | 0.31 | -0.03 | — | — | 92.99 | 39.06 |
| 29 | 12.70 | 0.82 | 0.19 | — | — | 55.22 | 39.52 |
| 30 | 13.72 | 0.96 | 0.32 | — | — | 76.75 | 39.80 |
| 31 | 12.78 | 0.78 | 0.08 | — | — | 69.79 | 39.86 |
| 32 | 14.72 | 1.16 | 0.49 | — | — | 83.14 | 40.27 |
| 33 | 10.48 | 1.70 | 0.58 | — | — | 73.46 | 40.96 |
| 34 | 14.76 | 0.81 | 0.26 | — | — | 94.99 | 40.97 |
| 35 | 14.65 | 0.73 | 0.11 | — | — | 57.69 | 41.13 |
| 36 | 12.61 | 0.86 | 0.16 | — | — | 60.85 | 41.59 |
| 37 | 15.36 | 0.45 | 0.04 | — | — | 53.14 | 41.73 |
| 38 | 15.46 | 0.43 | 0.14 | — | — | 82.84 | 42.84 |
| 39 | 15.22 | 0.50 | 0.63 | — | — | 86.89 | 42.88 |
| 40 | 11.43 | 0.97 | 0.22 | — | — | 63.85 | 43.16 |
| 41 | 14.79 | 0.38 | 0.06 | — | — | 46.63 | 45.32 |
| 42 | 14.80 | 0.49 | 0.18 | — | — | 89.16 | 45.71 |
| 43 | 11.90 | 1.00 | 0.19 | — | — | 31.66 | 46.18 |
| 44 | 14.35 | 0.49 | 0.02 | — | — | 82.32 | 46.19 |
| 45 | 11.48 | 1.06 | 0.22 | — | — | 61.04 | 48.18 |
| 46 | 14.71 | 0.56 | 0.26 | — | — | 70.36 | 48.32 |
| 47 | 15.57 | 0.04 | -0.01 | — | — | 77.60 | 48.42 |
| 48 | 14.78 | 0.85 | 0.50 | — | — | 67.24 | 48.51 |
| 49 | 12.49 | 0.90 | 0.18 | — | — | 64.57 | 49.14 |
| 50 | 13.89 | 0.56 | 0.05 | — | — | 73.33 | 49.67 |

Table 2. *(continued)*

| *Name* | *K* | *J − K* | *H − K* | *V − K* | *B − V* | *X* | *Y* |
|---|---|---|---|---|---|---|---|
| 51 | 15.40 | 0.41 | -0.04 | — | — | 81.50 | 49.95 |
| 52 | 15.69 | 0.27 | 0.04 | — | — | 86.68 | 50.61 |
| 53 | 12.80 | 0.91 | 0.14 | — | — | 51.07 | 50.64 |
| 54 | 14.15 | 0.80 | 0.26 | — | — | 29.90 | 53.31 |
| 55 | 14.32 | 0.50 | 0.05 | — | — | 20.60 | 55.17 |
| 56 | 14.15 | 0.16 | 0.40 | — | — | 40.95 | 56.41 |
| 57 | 11.57 | 1.04 | 0.20 | — | — | 30.26 | 56.91 |
| 58 | 12.22 | 0.91 | 0.23 | — | — | 37.71 | 56.97 |
| 59 | 15.46 | 0.70 | 0.45 | — | — | 52.54 | 60.23 |
| 60 | 12.77 | 1.08 | 0.25 | — | — | 44.43 | 61.78 |
| 61 | 15.46 | 0.46 | 0.08 | — | — | 56.49 | 64.71 |
| 62 | 14.77 | 0.43 | -0.05 | — | — | 65.67 | 65.41 |
| 63 | 15.48 | 0.39 | -0.15 | — | — | 66.46 | 67.92 |
| 64 | 15.76 | 0.19 | -0.41 | — | — | 63.11 | 68.18 |
| 65 | 15.17 | 0.54 | 0.14 | — | — | 47.12 | 72.32 |
| 66 | 12.16 | 0.95 | 0.20 | — | — | 76.74 | 72.69 |
| 67 | 15.44 | 0.54 | 0.02 | — | — | 62.23 | 75.47 |
| 68 | 14.18 | 0.66 | 0.12 | — | — | 27.32 | 76.95 |
| 69 | 14.97 | 0.99 | 0.27 | — | — | 55.13 | 92.45 |
| 70 | 14.99 | 0.60 | 0.02 | — | — | 70.77 | 98.18 |
| **NGC 1831** | | | | | | | |
| 1 | 14.15 | 0.94 | 0.32 | — | — | 4.78 | 6.57 |
| 2 | 15.69 | 0.14 | — | 1.82 | 0.83 | 94.80 | 7.08 |
| 3 | 14.87 | 0.62 | 0.07 | 3.05 | 1.18 | 70.84 | 9.02 |
| 4 | 15.43 | 0.13 | 0.30 | 1.16 | 0.14 | 79.36 | 12.27 |
| 5 | 14.82 | 0.64 | 0.13 | 2.71 | 1.14 | 55.16 | 17.70 |
| 6 | 15.20 | 0.70 | 0.29 | 2.92 | 1.36 | 90.16 | 18.70 |
| 7 | 15.51 | 0.47 | 0.14 | — | — | 2.73 | 18.95 |
| 8 | 14.74 | 0.45 | 0.12 | 1.99 | 0.70 | 104.42 | 19.53 |
| 9 | 15.86 | 0.29 | — | 2.06 | 0.17 | 94.63 | 19.61 |
| 10 | 15.34 | 0.46 | 0.16 | 1.93 | 0.79 | 68.74 | 20.46 |
| 11 | 15.75 | 0.34 | — | 1.80 | 0.67 | 42.90 | 20.56 |
| 12 | 15.63 | 0.06 | — | 2.08 | 0.47 | 96.25 | 21.55 |
| 13 | 15.72 | 0.29 | — | 2.85 | 0.58 | 94.29 | 22.50 |
| 14 | 14.57 | 0.65 | 0.17 | 2.80 | 1.20 | 48.25 | 22.74 |
| 15 | 15.79 | 0.22 | 0.07 | 1.93 | 0.85 | 79.19 | 23.77 |
| 16 | 15.76 | 0.41 | — | 2.02 | 0.60 | 101.26 | 24.58 |
| 17 | 14.87 | 0.76 | 0.13 | 2.96 | 1.11 | 102.72 | 27.15 |
| 18 | 14.26 | 0.69 | 0.19 | 2.84 | 0.74 | 97.80 | 28.19 |
| 19 | 15.92 | 0.42 | — | 2.14 | 1.17 | 102.99 | 31.98 |
| 20 | 11.67 | 1.04 | 0.25 | — | — | 10.88 | 33.65 |
| 21 | 13.71 | 0.68 | 0.08 | 3.32 | 1.50 | 95.39 | 35.08 |
| 22 | 15.47 | 0.53 | 0.27 | 2.33 | 0.89 | 56.26 | 35.10 |
| 23 | 14.50 | 0.66 | 0.22 | 2.86 | 1.09 | 103.07 | 38.18 |
| 24 | 15.43 | 0.51 | — | 2.36 | 1.08 | 92.27 | 38.67 |
| 25 | 14.88 | 1.07 | -0.36 | 2.95 | 0.23 | 90.55 | 39.41 |
| 26 | 10.10 | 1.97 | 0.75 | 6.90 | 4.23 | 71.19 | 40.11 |
| 27 | 11.69 | 1.15 | 0.27 | 4.33 | 1.70 | 78.00 | 41.50 |
| 28 | 15.33 | 0.65 | 0.28 | 2.31 | 0.83 | 71.37 | 81.83 |
| 29 | 12.47 | 0.92 | 0.19 | 3.69 | 1.60 | 86.29 | 84.21 |

**Table 2.** *(continued)*

| *Name* | $K$ | $J-K$ | $H-K$ | $V-K$ | $B-V$ | $X$ | $Y$ |
|---|---|---|---|---|---|---|---|
| **NGC 1868** | | | | | | | |
| 1 | 15.14 | -0.01 | -0.02 | — | — | 29.41 | 23.05 |
| 2 | 14.70 | 0.44 | -0.05 | 3.21 | 0.61 | 17.87 | 23.95 |
| 3 | 15.53 | 0.28 | 0.27 | — | — | 21.67 | 24.31 |
| 4 | 15.55 | 0.05 | — | — | — | 30.33 | 25.41 |
| 5 | 13.26 | 0.67 | 0.17 | 3.05 | — | 20.30 | 28.39 |
| 6 | 14.95 | 0.16 | 0.06 | 1.74 | — | 32.47 | 28.68 |
| 7 | 14.38 | 0.36 | 0.34 | — | — | 30.08 | 29.13 |
| 8 | 14.64 | 0.60 | 0.33 | — | — | 24.44 | 29.20 |
| 9 | 14.65 | 0.38 | 0.07 | 3.56 | 1.22 | 40.41 | 29.51 |
| 10 | 14.75 | 0.34 | 0.23 | — | — | 35.56 | 30.42 |
| 11 | 12.22 | 0.74 | 0.28 | 3.90 | — | 27.89 | 30.99 |
| 12 | 14.48 | 0.46 | 0.04 | — | — | 32.35 | 31.18 |
| 13 | 13.95 | 0.43 | 0.33 | 2.88 | — | 30.53 | 31.61 |
| 14 | 13.98 | 1.31 | -0.20 | 2.70 | — | 24.52 | 32.25 |
| 15 | 14.74 | 0.40 | 0.09 | — | — | 34.47 | 32.39 |
| 16 | 14.61 | 0.31 | 0.09 | — | — | 17.44 | 32.89 |
| 17 | 14.43 | 0.26 | 0.24 | — | — | 28.56 | 33.79 |
| 18 | 15.27 | 0.25 | 0.04 | — | — | 17.32 | 35.21 |
| 19 | 10.92 | 1.11 | 0.28 | 6.39 | 0.73 | 41.94 | 35.33 |
| 20 | 14.64 | 0.11 | -0.24 | — | — | 26.35 | 36.70 |
| 21 | 14.43 | 0.57 | 0.18 | 2.96 | 0.74 | 6.44 | 37.49 |
| 22 | 13.25 | 0.66 | 0.29 | 3.38 | — | 25.70 | 38.35 |
| 23 | 14.96 | 0.62 | 0.22 | — | — | 47.95 | 43.49 |
| 24 | 14.08 | 0.64 | 0.22 | — | — | 18.08 | 47.28 |
| 25 | 15.04 | 0.45 | -0.03 | — | — | 27.51 | 47.58 |
| **NGC 1978** | | | | | | | |
| 1 | 14.78 | 0.96 | 1.46 | — | — | 11.65 | 5.96 |
| 2 | 13.78 | 0.73 | 0.17 | — | — | 67.78 | 6.18 |
| 3 | 14.01 | 1.33 | -0.76 | — | — | 44.92 | 6.47 |
| 4 | 13.36 | 1.20 | 0.17 | — | — | 82.66 | 7.33 |
| 5 | 13.80 | 0.71 | 0.00 | — | — | 40.45 | 10.44 |
| 6 | 12.50 | 0.94 | 0.25 | — | — | 17.25 | 11.19 |
| 7 | 13.53 | 0.91 | 0.21 | — | — | 43.38 | 12.19 |
| 8 | 14.27 | 1.15 | 0.50 | — | — | 51.03 | 13.57 |
| 9 | 12.08 | 1.09 | 0.25 | — | — | 32.51 | 17.00 |
| 10 | 14.21 | 0.32 | 0.06 | — | — | 42.16 | 19.80 |
| 11 | 14.37 | 0.86 | 0.22 | — | — | 71.84 | 20.42 |
| 12 | 14.52 | 0.61 | -0.19 | — | — | 16.50 | 23.03 |
| 13 | 14.31 | 0.42 | -0.17 | — | — | 5.04 | 23.92 |
| 14 | 13.67 | 0.85 | 0.30 | — | — | 66.60 | 24.23 |
| 15 | 12.96 | 1.05 | 0.21 | — | — | 51.09 | 24.83 |
| 16 | 13.49 | 0.90 | 0.33 | — | — | 55.52 | 25.00 |
| 17 | 13.91 | 0.71 | 0.11 | — | — | 60.65 | 26.13 |
| 18 | 14.53 | 0.95 | 0.17 | — | — | 53.35 | 26.76 |
| 19 | 15.02 | 0.79 | 0.41 | — | — | 45.40 | 29.97 |
| 20 | 15.05 | 0.79 | 0.44 | — | — | 55.14 | 30.16 |
| 21 | 15.18 | 0.86 | 0.13 | — | — | 44.06 | 31.56 |
| 22 | 14.89 | 0.50 | 0.22 | — | — | 65.88 | 32.29 |
| 23 | 15.59 | -0.03 | — | — | — | 59.27 | 33.35 |
| 24 | 14.62 | 0.67 | 0.24 | — | — | 53.58 | 33.47 |
| 25 | 14.89 | 0.89 | -0.03 | — | — | 38.90 | 33.79 |

**Table 2.** *(continued)*

| *Name* | $K$ | $J-K$ | $H-K$ | $V-K$ | $B-V$ | $X$ | $Y$ |
|---|---|---|---|---|---|---|---|
| 26 | 15.14 | 0.28 | -0.42 | — | — | 51.98 | 33.80 |
| 27 | 13.51 | 0.93 | 0.19 | — | — | 68.91 | 34.29 |
| 28 | 15.42 | 0.61 | -0.01 | — | — | 48.22 | 34.97 |
| 29 | 14.61 | 1.07 | -0.01 | — | — | 26.31 | 36.90 |
| 30 | 13.52 | 0.74 | 0.39 | — | — | 74.40 | 36.98 |
| 31 | 14.15 | 0.74 | 0.16 | — | — | 39.58 | 37.85 |
| 32 | 14.27 | 0.89 | 0.22 | — | — | 52.94 | 37.96 |
| 33 | 12.67 | 0.90 | 0.27 | — | — | 49.73 | 38.10 |
| 34 | 14.56 | 0.92 | 0.32 | — | — | 46.82 | 38.11 |
| 35 | 14.18 | 0.65 | 0.11 | — | — | 55.81 | 38.62 |
| 36 | 15.10 | 0.89 | 0.04 | — | — | 38.25 | 40.35 |
| 37 | 15.24 | 0.68 | 0.18 | — | — | 43.77 | 40.55 |
| 38 | 14.72 | 0.92 | 0.17 | — | — | 52.94 | 40.71 |
| 39 | 14.18 | 0.48 | 0.30 | — | — | 68.40 | 41.46 |
| 40 | 14.72 | 0.56 | 0.17 | — | — | 85.04 | 41.62 |
| 41 | 14.80 | 0.41 | 0.23 | — | — | 65.90 | 41.81 |
| 42 | 14.93 | 0.80 | 0.04 | — | — | 87.54 | 42.16 |
| 43 | 15.04 | 0.69 | 0.35 | — | — | 55.95 | 42.17 |
| 44 | 14.49 | 0.65 | 0.48 | — | — | 52.17 | 43.13 |
| 45 | 13.67 | 0.73 | 0.18 | — | — | 35.18 | 43.49 |
| 46 | 14.72 | 0.39 | 0.38 | — | — | 62.05 | 43.78 |
| 47 | 15.17 | 0.07 | -0.01 | — | — | 58.34 | 43.89 |
| 48 | 12.78 | 0.68 | 0.40 | — | — | 49.93 | 44.18 |
| 49 | 15.21 | 0.14 | -0.14 | — | — | 56.01 | 44.83 |
| 50 | 14.65 | 0.42 | -0.54 | — | — | 53.37 | 44.97 |
| 51 | 14.61 | 0.69 | 0.25 | — | — | 45.85 | 45.00 |
| 52 | 14.85 | 0.75 | 0.36 | — | — | 76.24 | 45.52 |
| 53 | 13.69 | 1.70 | 1.10 | — | — | 41.59 | 46.46 |
| 54 | 14.91 | 0.76 | — | — | — | 51.39 | 47.02 |
| 55 | 14.05 | 0.12 | 0.07 | — | — | 57.42 | 47.08 |
| 56 | 13.87 | 0.81 | 0.44 | — | — | 36.72 | 47.45 |
| 57 | 14.39 | 0.59 | 0.55 | — | — | 43.36 | 47.71 |
| 58 | 15.01 | -0.40 | — | — | — | 42.22 | 47.75 |
| 59 | 14.97 | 0.09 | -0.23 | — | — | 78.39 | 47.82 |
| 60 | 14.12 | 0.96 | 0.07 | — | — | 72.12 | 48.25 |
| 61 | 13.66 | 1.47 | 0.44 | — | — | 74.79 | 48.32 |
| 62 | 11.22 | 1.27 | 0.58 | — | — | 65.43 | 48.48 |
| 63 | 13.89 | 0.66 | 0.19 | — | — | 47.52 | 48.90 |
| 64 | 13.85 | 0.53 | — | — | — | 69.05 | 49.08 |
| 65 | 14.30 | 1.08 | -0.38 | — | — | 62.89 | 49.64 |
| 66 | 14.16 | 0.71 | 0.04 | — | — | 35.44 | 49.83 |
| 67 | 13.34 | 0.80 | 0.15 | — | — | 55.49 | 50.02 |
| 68 | 13.10 | 0.71 | 0.26 | — | — | 51.54 | 50.11 |
| 69 | 14.66 | 0.45 | -0.11 | — | — | 40.37 | 50.15 |
| 70 | 12.77 | 1.37 | 0.86 | — | — | 57.79 | 50.51 |
| 71 | 13.26 | 1.06 | 0.55 | — | — | 60.79 | 51.39 |
| 72 | 12.86 | 0.79 | 0.29 | — | — | 74.32 | 51.83 |
| 73 | 15.52 | 0.17 | 0.24 | — | — | 31.53 | 52.44 |
| 74 | 15.13 | 0.41 | -0.36 | — | — | 34.29 | 52.56 |
| 75 | 14.82 | 1.15 | 1.68 | — | — | 47.19 | 53.86 |
| 76 | 13.99 | 0.55 | — | — | — | 52.27 | 54.39 |
| 77 | 15.72 | -0.12 | 0.90 | — | — | 28.53 | 54.91 |
| 78 | 9.83 | 2.03 | — | — | — | 59.83 | 54.94 |

**Table 2.** *(continued)*

| *Name* | $K$ | $J-K$ | $H-K$ | $V-K$ | $B-V$ | $X$ | $Y$ |
|---|---|---|---|---|---|---|---|
| 79 | 11.24 | 1.20 | — | — | — | 65.16 | 55.08 |
| 80 | 12.23 | 1.89 | — | — | — | 62.02 | 55.41 |
| 81 | 16.54 | 0.09 | -0.27 | — | — | 7.22 | 55.60 |
| 82 | 13.70 | 0.94 | — | — | — | 69.76 | 55.87 |
| 83 | 16.01 | 0.05 | -0.98 | — | — | 4.20 | 56.01 |
| 84 | 16.06 | -0.83 | — | — | — | 1.56 | 56.03 |
| 85 | 14.20 | 0.24 | — | — | — | 66.09 | 57.71 |
| 86 | 13.19 | 0.37 | -0.22 | — | — | 34.30 | 57.79 |
| 87 | 13.26 | 0.84 | — | — | — | 57.74 | 58.29 |
| 88 | 15.32 | 0.25 | 0.95 | — | — | 41.46 | 58.53 |
| 89 | 12.03 | 0.87 | — | — | — | 60.42 | 59.49 |
| 90 | 13.73 | 0.58 | 0.16 | — | — | 48.01 | 59.77 |
| 91 | 14.88 | -0.08 | 0.17 | — | — | 51.67 | 59.80 |
| 92 | 13.16 | 0.79 | 0.29 | — | — | 76.70 | 62.08 |
| 93 | 15.60 | -0.03 | 0.16 | — | — | 20.01 | 62.20 |
| 94 | 14.02 | 0.83 | 0.52 | — | — | 89.95 | 62.56 |
| 95 | 13.81 | 0.47 | -0.12 | — | — | 68.91 | 62.82 |
| 96 | 13.59 | 0.73 | 0.25 | — | — | 23.52 | 63.42 |
| 97 | 14.54 | 0.76 | 0.05 | — | — | 54.05 | 63.45 |
| 98 | 10.86 | 1.33 | 0.37 | — | — | 48.23 | 63.49 |
| 99 | 16.50 | -0.66 | -0.09 | — | — | 72.51 | 63.90 |
| 100 | 13.01 | 0.78 | 0.20 | — | — | 59.03 | 64.77 |
| 101 | 10.86 | — | 1.67 | — | — | 89.12 | 65.22 |
| 102 | 15.10 | 0.30 | 0.34 | — | — | 40.77 | 66.11 |
| 103 | 11.50 | 1.03 | 0.25 | — | — | 65.86 | 66.96 |
| 104 | 15.74 | -0.09 | 0.57 | — | — | 74.45 | 67.17 |
| 105 | 13.11 | 0.91 | 0.21 | — | — | 54.14 | 67.20 |
| 106 | 13.65 | 0.72 | 0.25 | — | — | 57.63 | 67.32 |
| 107 | 15.27 | 1.05 | 0.63 | — | — | 46.60 | 67.85 |
| 108 | 15.62 | 0.58 | 0.11 | — | — | 57.69 | 71.03 |
| 109 | 13.43 | 0.70 | 0.22 | — | — | 64.48 | 72.22 |
| 110 | 14.81 | 0.87 | 1.03 | — | — | 40.09 | 72.26 |
| 111 | 14.08 | 0.50 | 0.33 | — | — | 50.23 | 72.61 |
| 112 | 15.86 | -0.17 | -0.16 | — | — | 59.90 | 72.93 |
| 113 | 15.03 | 0.45 | — | — | — | 53.39 | 74.35 |
| 114 | 12.67 | 0.94 | 0.27 | — | — | 77.04 | 74.81 |
| 115 | 14.69 | 0.50 | 0.14 | — | — | 67.49 | 75.42 |
| 116 | 14.69 | 0.36 | 0.24 | — | — | 58.98 | 76.32 |
| 117 | 14.80 | 0.67 | 0.63 | — | — | 30.78 | 77.78 |
| 118 | 13.73 | 0.65 | 0.21 | — | — | 81.28 | 79.76 |
| 119 | 13.24 | 0.91 | 0.22 | — | — | 30.27 | 82.38 |
| 120 | 15.12 | 0.09 | 0.06 | — | — | 94.01 | 85.40 |
| 121 | 13.53 | 0.77 | 0.25 | — | — | 57.26 | 90.26 |
| 122 | 14.57 | 0.56 | 0.70 | — | — | 37.97 | 90.60 |
| 123 | 14.60 | 0.61 | 0.44 | — | — | 8.30 | 92.75 |
| 124 | 11.23 | 1.46 | 0.41 | — | — | 49.10 | 98.29 |
| **NGC 1987** | | | | | | | |
| 1 | 11.55 | 1.11 | 0.26 | 5.09 | 1.72 | 47.69 | 9.23 |
| 2 | 12.26 | 0.89 | 0.24 | 3.89 | 1.57 | 50.24 | 13.82 |
| 3 | 14.16 | 0.66 | 0.25 | 2.81 | 1.19 | 25.68 | 20.43 |
| 4 | 15.17 | 0.58 | — | 2.63 | 0.87 | 65.97 | 29.34 |
| 5 | 14.99 | 0.68 | 0.45 | 2.68 | 1.15 | 60.60 | 29.76 |

**Table 2.** *(continued)*

| *Name* | $K$ | $J-K$ | $H-K$ | $V-K$ | $B-V$ | $X$ | $Y$ |
|---|---|---|---|---|---|---|---|
| 6 | 15.50 | 0.32 | 0.08 | 2.02 | 0.56 | 59.78 | 35.35 |
| 7 | 10.09 | 1.96 | 0.76 | 6.67 | 3.84 | 27.84 | 40.33 |
| 8 | 10.87 | 1.14 | 0.28 | 5.36 | 1.42 | 57.48 | 43.01 |
| 9 | 11.35 | 1.00 | 0.32 | 4.72 | 1.73 | 57.91 | 45.68 |
| 10 | 13.08 | 0.84 | 0.20 | 3.30 | 1.20 | 54.54 | 47.43 |
| 11 | 13.22 | 0.82 | 0.19 | 3.36 | 1.34 | 67.20 | 47.98 |
| 12 | 13.73 | 0.90 | 0.17 | 3.62 | 1.62 | 20.72 | 53.02 |
| 13 | 13.10 | 0.85 | 0.19 | 3.64 | 1.58 | 48.09 | 56.37 |
| 14 | 15.22 | 0.34 | 0.27 | 2.50 | 0.95 | 42.02 | 63.55 |
| 15 | 14.27 | 0.61 | 0.09 | 2.87 | 1.11 | 45.14 | 65.28 |
| 16 | 15.69 | 0.19 | -0.10 | 2.30 | 1.09 | 50.09 | 65.43 |
| 17 | 15.82 | 0.58 | 0.28 | 2.40 | 1.11 | 40.80 | 67.39 |
| 18 | 13.34 | 0.87 | 0.26 | 3.59 | 1.54 | 66.10 | 75.33 |
| 19 | 11.94 | 1.10 | 0.27 | 4.88 | 1.84 | 31.97 | 75.96 |
| 20 | 13.51 | 0.75 | 0.18 | 3.49 | 1.47 | 17.86 | 82.21 |
| 21 | 14.39 | 0.74 | 0.47 | 2.95 | 1.32 | 83.09 | 86.86 |
| 22 | 10.90 | 1.08 | 0.31 | 5.53 | 1.84 | 91.50 | 96.82 |
| 23 | 13.81 | 0.79 | 0.32 | — | — | 1.01 | 100.19 |
| **NGC 2107** | | | | | | | |
| 1 | 10.14 | 1.11 | 0.22 | — | — | 20.78 | 15.76 |
| 2 | 13.36 | 1.62 | 0.54 | — | — | 25.60 | 15.78 |
| 3 | 14.21 | 0.79 | 0.16 | — | — | 52.97 | 24.54 |
| 4 | 12.76 | 0.98 | 0.20 | — | — | 9.76 | 35.49 |
| 5 | 14.47 | 0.65 | 0.29 | — | — | 42.99 | 35.58 |
| 6 | 14.40 | 1.12 | 0.42 | — | — | 53.16 | 36.46 |
| 7 | 14.73 | 0.62 | 0.22 | — | — | 37.87 | 36.92 |
| **NGC 2108** | | | | | | | |
| 1 | 15.73 | 0.25 | 0.09 | — | — | 53.29 | 25.15 |
| 2 | 15.44 | 0.22 | 0.33 | 0.75 | -0.03 | 85.28 | 28.63 |
| 3 | 14.25 | 0.94 | 0.59 | 3.28 | 0.91 | 65.85 | 37.53 |
| 4 | 10.88 | 1.59 | -0.29 | 6.19 | 1.64 | 51.60 | 40.53 |
| 5 | 12.45 | 0.81 | 0.40 | 4.06 | 1.61 | 66.66 | 40.69 |
| 6 | 12.64 | 1.59 | 0.31 | 4.38 | 1.14 | 68.39 | 41.36 |
| 7 | 14.83 | 0.69 | 0.54 | 2.59 | 0.72 | 66.19 | 45.74 |
| 8 | 13.28 | 0.94 | 0.35 | 4.66 | 0.12 | 39.70 | 46.16 |
| 9 | 16.00 | 0.04 | 0.01 | 1.90 | 1.09 | 77.64 | 48.73 |
| 10 | 15.27 | 0.68 | 0.47 | — | — | 44.68 | 50.25 |
| 11 | 13.65 | 1.04 | 0.51 | 3.47 | 1.25 | 46.21 | 53.88 |
| 12 | 14.94 | 0.91 | 0.66 | — | — | 55.73 | 54.05 |
| 13 | 13.17 | 1.09 | 0.56 | 3.73 | 1.39 | 60.91 | 56.21 |
| 14 | 13.92 | 0.86 | 0.58 | 3.16 | 1.04 | 69.18 | 57.29 |
| 15 | 14.02 | 1.06 | 0.84 | 3.53 | 1.23 | 2.84 | 58.91 |
| 16 | 12.84 | 1.26 | 0.45 | 4.27 | 1.68 | 81.27 | 60.52 |
| 17 | 14.10 | 1.22 | 0.35 | 3.67 | 1.44 | 69.22 | 60.90 |
| 18 | 13.84 | 1.19 | 0.53 | 3.66 | 1.34 | 54.55 | 61.53 |
| 19 | 14.07 | 1.14 | 0.91 | 3.37 | 1.20 | 5.93 | 75.60 |
| 20 | 14.29 | 1.15 | 0.52 | — | — | 39.32 | 91.45 |

**Table 2.** *(continued)*

| *Name* | *K* | *J − K* | *H − K* | *V − K* | *B − V* | *X* | *Y* |
|---|---|---|---|---|---|---|---|
| **NGC 2162** | | | | | | | |
| 1 | 14.48 | 0.64 | -0.04 | 2.93 | 1.22 | 31.07 | 6.33 |
| 2 | 17.12 | -0.28 | -0.63 | 1.93 | 0.91 | 29.21 | 8.07 |
| 3 | 14.88 | 0.84 | 0.25 | 3.01 | 1.12 | 59.60 | 10.98 |
| 4 | 16.05 | 1.41 | — | 3.15 | 0.81 | 60.61 | 12.39 |
| 5 | 16.55 | 1.38 | — | 2.57 | 0.87 | 57.74 | 14.67 |
| 6 | 15.18 | 0.60 | 0.01 | 2.71 | 1.06 | 58.22 | 14.95 |
| 7 | 16.47 | 0.77 | -0.50 | 2.51 | 0.88 | 56.26 | 15.34 |
| 8 | 15.85 | 0.64 | 0.13 | 2.23 | 0.87 | 45.26 | 15.77 |
| 9 | 15.81 | 0.81 | 0.10 | 2.79 | 0.74 | 43.51 | 15.87 |
| 10 | 15.78 | 0.68 | 0.00 | 2.10 | 1.05 | 59.55 | 16.90 |
| 11 | 16.05 | 0.30 | 0.09 | 3.57 | 0.73 | 58.77 | 17.35 |
| 12 | 15.71 | 0.89 | 0.57 | 2.95 | 0.68 | 44.54 | 17.54 |
| 13 | 15.66 | 0.78 | -0.41 | 2.68 | 0.89 | 30.16 | 18.50 |
| 14 | 15.86 | 1.06 | 0.21 | 3.44 | 0.56 | 37.30 | 18.60 |
| 15 | 14.51 | 0.58 | 0.05 | 2.70 | 1.20 | 36.13 | 18.89 |
| 16 | 16.41 | 0.34 | 1.04 | 2.01 | 0.76 | 29.56 | 19.39 |
| 17 | 15.42 | 0.62 | 0.23 | 2.42 | 1.00 | 53.53 | 20.31 |
| 18 | 16.67 | 0.46 | 0.33 | 1.59 | 0.54 | 45.74 | 20.55 |
| 19 | 16.02 | 0.17 | -0.30 | 2.27 | 0.98 | 49.57 | 21.23 |
| 20 | 17.14 | -0.61 | -0.77 | 0.87 | 0.12 | 37.52 | 21.30 |
| 21 | 16.53 | 0.93 | -0.39 | 2.22 | 0.79 | 51.79 | 21.39 |
| 22 | 15.23 | 0.39 | -0.07 | 2.85 | 1.18 | 34.19 | 23.13 |
| 23 | 16.44 | — | 0.16 | 3.32 | 0.77 | 48.49 | 23.15 |
| 24 | 13.05 | 0.85 | 0.17 | 3.48 | 1.41 | 31.51 | 23.45 |
| 25 | 13.40 | 0.73 | 0.28 | 3.22 | 1.40 | 45.49 | 24.06 |
| 26 | 14.97 | 1.35 | -0.35 | 3.89 | 0.90 | 46.41 | 24.11 |
| 27 | 16.59 | 1.08 | 0.92 | 2.46 | 0.87 | 55.74 | 25.57 |
| 28 | 13.40 | 0.83 | 0.22 | 4.46 | 1.62 | 27.84 | 26.30 |
| 29 | 15.99 | 0.51 | -0.35 | 2.42 | 1.01 | 55.03 | 26.95 |
| 30 | 16.44 | 1.29 | -0.02 | 2.63 | 0.64 | 32.04 | 27.81 |
| 31 | 16.68 | 0.07 | 0.70 | 2.45 | 0.85 | 54.26 | 28.27 |
| 32 | 14.87 | 0.87 | 0.00 | 2.91 | 1.25 | 49.89 | 28.27 |
| 33 | 15.61 | 0.53 | 0.16 | 2.68 | 1.02 | 32.99 | 28.89 |
| 34 | 15.46 | 1.23 | 0.68 | 3.47 | 0.67 | 46.55 | 30.39 |
| 35 | 15.84 | 0.56 | -0.23 | 2.60 | 0.62 | 50.35 | 30.62 |
| 36 | 11.54 | 1.00 | 0.17 | 4.55 | 1.80 | 41.23 | 30.62 |
| 37 | 14.11 | 0.63 | 0.02 | 3.09 | 1.35 | 50.73 | 40.76 |
| 38 | 15.87 | 1.61 | 0.06 | 3.16 | 0.90 | 32.85 | 41.46 |
| 39 | 16.15 | 0.11 | 0.09 | 2.16 | 1.09 | 33.87 | 41.86 |
| 40 | 11.53 | 1.09 | 0.18 | 4.56 | 1.75 | 35.76 | 45.14 |
| 41 | 15.20 | 0.46 | -0.08 | 2.43 | 1.02 | 33.43 | 47.80 |
| 42 | 15.23 | 0.52 | -0.07 | 2.65 | 1.22 | 54.43 | 48.22 |
| **NGC 2173** | | | | | | | |
| 1 | 13.61 | 0.93 | 0.03 | 3.58 | 1.52 | 42.31 | 5.69 |
| 2 | 14.06 | 0.91 | 0.34 | 3.55 | 1.28 | 58.17 | 6.45 |
| 3 | 14.00 | 1.05 | 0.52 | 3.54 | 1.42 | 60.19 | 10.74 |
| 4 | 14.96 | 0.87 | -0.05 | 3.15 | 1.27 | 58.51 | 19.37 |
| 5 | 14.59 | 0.66 | -0.09 | 2.89 | 1.28 | 28.28 | 27.61 |

**Table 2.** *(continued)*

| *Name* | $K$ | $J-K$ | $H-K$ | $V-K$ | $B-V$ | $X$ | $Y$ |
|---|---|---|---|---|---|---|---|
| 6 | 11.06 | 1.32 | 0.40 | 5.33 | 1.78 | 60.89 | 27.84 |
| 7 | 13.04 | 1.08 | 0.32 | — | — | 84.46 | 33.36 |
| 8 | 13.87 | 0.88 | 0.28 | — | — | 94.54 | 36.69 |
| 9 | 11.00 | 1.60 | 0.57 | 5.34 | 2.22 | 58.77 | 38.56 |
| 10 | 11.15 | 1.18 | 0.33 | 4.80 | 1.68 | 55.87 | 40.26 |
| 11 | 13.54 | 0.95 | 0.24 | — | — | 99.92 | 44.29 |
| 12 | 13.81 | 1.08 | 0.40 | 3.63 | 1.42 | 56.34 | 45.35 |
| 13 | 12.72 | 0.95 | 0.29 | 3.58 | 1.42 | 68.37 | 45.61 |
| 14 | 13.26 | 0.92 | 0.23 | 3.50 | 1.27 | 71.47 | 52.00 |
| 15 | 14.05 | 0.30 | 0.12 | 2.01 | 0.81 | 52.42 | 56.51 |
| 16 | 12.61 | 0.90 | 0.06 | 4.11 | 1.81 | 7.76 | 57.79 |
| 17 | 13.95 | 0.76 | 0.20 | 3.42 | 1.42 | 74.48 | 58.60 |
| 18 | 12.16 | 1.00 | 0.20 | 3.93 | 1.66 | 79.83 | 59.07 |
| 19 | 13.78 | 0.77 | 0.17 | 3.36 | 1.39 | 41.57 | 64.47 |
| 20 | 15.70 | 0.33 | -0.06 | — | — | 14.14 | 64.84 |
| 21 | 15.56 | 0.56 | 0.10 | 2.35 | 1.14 | 14.14 | 64.84 |
| 22 | 13.73 | 0.80 | 0.12 | 3.56 | 1.58 | 75.64 | 78.30 |
| 23 | 14.95 | 0.93 | 0.62 | 3.11 | 1.24 | 30.45 | 82.43 |
| 24 | 15.57 | 0.69 | 0.43 | 2.97 | 1.18 | 50.96 | 94.41 |
| 25 | 12.19 | 1.03 | 0.24 | 4.17 | 1.75 | 74.77 | 98.45 |
| **NGC 2209** | | | | | | | |
| 1 | 15.59 | 0.33 | — | 2.32 | 1.15 | 29.02 | 4.10 |
| 2 | 16.13 | 0.75 | — | 2.52 | 1.11 | 24.54 | 20.30 |
| 3 | 10.39 | 1.90 | 0.69 | 6.16 | 2.57 | 84.57 | 22.72 |
| 4 | 14.80 | 0.98 | 0.40 | 3.06 | 1.07 | 54.68 | 23.57 |
| 5 | 11.88 | 1.08 | 0.18 | 4.46 | 1.79 | 19.09 | 35.53 |
| 6 | 14.87 | 1.14 | 0.36 | 3.30 | 1.26 | 72.06 | 42.00 |
| 7 | 14.74 | 1.20 | 0.47 | 3.36 | 1.22 | 60.20 | 44.74 |
| 8 | 15.14 | 0.89 | 0.40 | 3.20 | 1.18 | 71.90 | 48.74 |
| 9 | 15.30 | 0.24 | 0.09 | 1.95 | 0.93 | 52.65 | 64.68 |
| 10 | 15.32 | 0.64 | -0.03 | 2.83 | 1.21 | 9.34 | 97.77 |

**Table 3.** Cross-identifications and comparisons with the literature.

| Cluster | Name | MA source | $K$ | $J-K$ | $H-K$ | FMB | $K$ | $J-K$ | $H-K$ | this paper | $K$ | $J-K$ | $H-K$ |
|---|---|---|---|---|---|---|---|---|---|---|---|---|---|
| NGC 1756 | LE1 | [4] | 11.74 | 0.85 | 0.22 | – | – | – | – | 25 | 11.73 | 0.81 | 0.26 |
| NGC 1783 | LE2 | [3] | 11.26 | 1.07 | 0.23 | – | – | – | – | 1 | 11.37 | 1.15 | 0.26 |
| | LE3 | – | – | – | – | 15 | 10.36 | 1.60 | 0.59 | 71 | 10.61 | 1.89 | 0.73 |
| | LE5 | [3] | 11.23 | 1.08 | 0.23 | 14 | 11.23 | 1.09 | 0.23 | 35 | 11.38 | 1.03 | 0.30 |
| | LE6 | – | – | – | – | 4 | 11.34 | 1.08 | 0.22 | 19 | 11.39 | 1.17 | 0.20 |
| | LE9 | [2] | 10.97 | 1.10 | 0.26 | 13 | 10.93 | 1.11 | 0.27 | 52 | 11.04 | 1.12 | 0.22 |
| | LE10 | [2] | 11.40 | 1.07 | 0.23 | 7 | 11.36 | 1.06 | 0.21 | 37 | 11.43 | 1.07 | 0.22 |
| | LE11 | [3] | 11.55 | 1.04 | 0.17 | 11 | 11.51 | 1.00 | 0.18 | 53 | 11.65 | 1.08 | 0.19 |
| | – | – | – | – | – | 12 | 11.84 | 0.97 | 0.18 | 60 | 12.13 | 0.99 | 0.20 |
| NGC 1806 | LE1 | [2] | 10.34 | 1.60 | 0.59 | 3 | 10.42 | 1.68 | 0.63 | 33 | 10.48 | 1.70 | 0.58 |
| | LE2 | [2] | 11.36 | 1.02 | 0.21 | 4 | 11.28 | 1.04 | 0.21 | 40 | 11.43 | 0.97 | 0.22 |
| | LE3 | [2] | 11.45 | 1.06 | 0.20 | 5 | 11.37 | 1.02 | 0.21 | 45 | 11.48 | 1.06 | 0.22 |
| | LE4 | [3] | 11.59 | 1.04 | 0.20 | – | 11.58 | 1.00 | 0.18 | 19 | 11.69 | 0.95 | 0.19 |
| | LE5 | [3] | 11.05 | 1.03 | 0.19 | 2 | 11.08 | 1.04 | 0.19 | 2 | 11.23 | 1.08 | 0.19 |
| | LE6 | [4] | 11.48 | 1.09 | 0.22 | 6 | 11.41 | 1.07 | 0.22 | 57 | 11.57 | 1.04 | 0.20 |
| | LE7 | [4] | 12.09 | 1.03 | 0.16 | 7 | 12.04 | 1.03 | 0.19 | 66 | 12.16 | 0.95 | 0.20 |
| | LE8 | [4] | 11.82 | 1.03 | 0.20 | 8 | 11.79 | 0.99 | 0.18 | 43 | 11.90 | 1.00 | 0.19 |
| | MA9 | [2] | 10.35 | 1.80 | 0.70 | 1 | 10.32 | 1.81 | 0.68 | 1 | 10.43 | 1.91 | 0.70 |
| NGC 1831 | MA1 | [2] | 10.22 | 1.83 | 0.74 | – | – | – | – | 26 | 10.10 | 1.97 | 0.75 |
| | MA2 | [2] | 11.73 | 1.01 | 0.20 | – | – | – | – | 27 | 11.69 | 1.15 | 0.27 |
| NGC 1978 | LE3 | [2] | 9.90 | 1.81 | 0.73 | – | – | – | – | 78 | 9.83 | 2.03 | – |
| | LE4 | [2] | 11.23 | 1.05 | 0.23 | 4 | 11.29 | 1.13 | 0.25 | 79 | 11.24 | 1.20 | – |
| | LE5 | [4] | 11.44 | 1.05 | 0.22 | 3 | 11.50 | 1.13 | 0.25 | 103 | 11.50 | 1.03 | 0.25 |
| | LE6 | [3] | 10.59 | 1.23 | 0.37 | 9 | 10.83 | 1.35 | 0.39 | 98 | 10.86 | 1.33 | 0.37 |
| | LE9 | [3] | 12.36 | 1.06 | 0.20 | 15 | 12.36 | 1.06 | 0.20 | 9 | 12.08 | 1.09 | 0.25 |
| NGC 1987 | LE4 | [2] | 10.91 | 1.13 | 0.27 | 7 | 10.83 | 1.14 | 0.27 | 22 | 10.90 | 1.16 | 0.31 |
| | LE5 | [2] | 10.21 | 1.70 | 0.65 | 2 | 10.16 | 1.73 | 0.63 | 7 | 10.09 | 1.96 | 0.77 |
| | LE6 | – | – | – | – | 4 | 11.53 | 1.10 | 0.24 | 1 | 11.55 | 1.08 | 0.26 |
| | – | – | – | – | – | 1 | 11.98 | 1.07 | 0.21 | 19 | 11.94 | 1.10 | 0.27 |
| | – | – | – | – | – | 3 | 12.24 | 0.98 | 0.18 | 2 | 12.26 | 0.89 | 0.24 |

**Table 4.** Observed and normalized numbers of AGB and RGB stars.

| *Name* | $N_{AGB}$ | $N_{RGB}$ | $m_{Bol}$ | $L/L_{\odot}$ | $N4$ | $N^*_{AGB}$ | $N^*_{RGB}$ | $s$ |
|---|---|---|---|---|---|---|---|---|
| NGC 1756 | 1 | 1 | 11.66 | 4.74 | 0.84 | 0.21 | 0.21 | 32 |
| NGC 1783 | 8 | 26 | 10.50 | 13.80 | 2.82 | 0.58 | 1.81 | 37 |
| NGC 1806 | 10 | 22 | 10.37 | 15.56 | 2.38 | 0.64 | 1.41 | 40 |
| NGC 1831 | 3 | 2 | 10.41 | 15.00 | 0.47 | 0.20 | 0.13 | 31 |
| NGC 1868 | 2 | 1 | 11.15 | 7.59 | 0.79 | 0.26 | 0.13 | 33 |
| NGC 1978 | 8 | 38 | 10.00 | 21.88 | 2.47 | 0.37 | 1.69 | 45 |
| NGC 1987 | 7 | 7 | 11.14 | 7.66 | 2.09 | 0.91 | 0.91 | 35 |
| NGC 2107 | 1 | 0 | 10.85 | 10.00 | 0.30 | 0.10 | 0.00 | 32 |
| NGC 2108 | 1 | 10 | 11.30 | 6.61 | 1.97 | 0.15 | 1.51 | 36 |
| NGC 2162 | 2 | 3 | 12.60 | 1.99 | 3.00 | 1.00 | 1.50 | 39 |
| NGC 2173 | 5 | 10 | 11.46 | 5.70 | 3.16 | 0.88 | 1.75 | 42 |
| NGC 2209 | 2 | 0 | 12.75 | 1.74 | 1.15 | 1.15 | 0.00 | 35 |

**Table 5.** AGB and RGB contribution to the luminosity of the cluster.

| *Name* | $L_K^{AGB}/L_T$ | $L_K^{RGB}/L_T$ | $m_{Bol}$ | $L_{Bol}^{RGB}/L_T$ | $s$ |
|---|---|---|---|---|---|
| NGC 1756 | – | – | 11.66 | 0.00 | 32 |
| NGC 1783 | 0.64 | 0.23 | 10.50 | 0.16 | 37 |
| NGC 1806 | 0.58 | 0.21 | 10.37 | 0.19 | 40 |
| NGC 1831 | 0.65 | 0.04 | 10.41 | 0.01 | 31 |
| NGC 1868 | 0.43 | 0.02 | 11.15 | 0.02 | 33 |
| NGC 1978 | 0.48 | 0.22 | 10.00 | 0.19 | 45 |
| NGC 1987 | 1.05 | 0.11 | 11.14 | 0.10 | 35 |
| NGC 2107 | 0.42 | 0.00 | 10.85 | 0.00 | 32 |
| NGC 2108 | 0.23 | 0.18 | 11.30 | 0.11 | 36 |
| NGC 2162 | 0.63 | 0.14 | 12.60 | 0.16 | 39 |
| NGC 2173 | 0.60 | 0.20 | 11.46 | 0.18 | 42 |
| NGC 2209 | 0.91 | 0.00 | 12.75 | 0.00 | 35 |

**Table 6.** Theoretical quantities.

| SGR | | Al93C | | Al93O | |
|---|---|---|---|---|---|
| *Logt* | $L_{bol}^{RGB}/L_T$ | *Logt* | $L_{bol}^{RGB}/L_T$ | *Logt* | $L_{bol}^{RGB}/L_T$ |
| 9.300 | 0.17 | 9.471 | 0.17 | 9.504 | 0.16 |
| 9.123 | 0.17 | 9.351 | 0.17 | 9.406 | 0.16 |
| 8.982 | 0.15 | 9.252 | 0.16 | 9.314 | 0.16 |
| 8.864 | 0.12 | 9.172 | 0.16 | 9.281 | 0.09 |
| 8.810 | 0.08 | 9.094 | 0.15 | 9.203 | 0.04 |
| 8.784 | 0.06 | 9.024 | 0.14 | 9.131 | - |
| 8.761 | 0.03 | 8.957 | 0.12 | 9.065 | - |
| 8.736 | - | 8.894 | 0.08 | 9.001 | - |
| | | 8.835 | - | 8.536 | - |

SGR = Sweigart, Greggio and Renzini (1989,1990) *standard models*
Al93C = Alongi *et al* (1993) *standard models*
Al93O = Alongi *et al* (1993) *overshooting models*

**Table 7.** Comparisons between observed and computed integrated colours and magnitudes.

| *Name* | Aperture | $K$ | $J-K$ | $K_{our}$ | $(J-K)_{our}$ |
|---|---|---|---|---|---|
| NGC 1756 | – | – | – | 10.56 | 0.62 |
| NGC 1783 | 60 | 8.56 | 0.74 | 8.50 | 1.02 |
| NGC 1806 | 60 | 8.19 | 0.89 | 8.33 | 1.07 |
| NGC 1831 | 59 | 9.21 | 0.48 | 9.43 | 1.33 |
| NGC 1868 | 64 | 9.73 | 0.69 | 10.05 | 0.76 |
| NGC 1978 | 60 | 7.92 | 0.93 | 7.96 | 1.14 |
| NGC 1987 | 60 | 9.01 | 0.89 | 8.79 | 1.22 |
| NGC 2107 | 60 | 9.21 | 0.80 | 9.93 | 1.09 |
| NGC 2108 | 64 | 9.27 | 1.18 | 9.94 | 1.24 |
| NGC 2162 | 30 | 10.39 | 0.95 | 10.17 | 0.89 |
| NGC 2173 | 30 | 9.90 | 1.04 | 9.24 | 1.14 |
| NGC 2209 | 30 | 10.04 | 1.68 | 10.06 | 1.58 |

**Table 8.** AGB and RGB contribution to the integrated colours.

| *Name* | $(J-K)$ | $(J-K)_{AGB}$ | $(J-K)_{RGB}$ | $(V-K)$ | $(V-K)_{AGB}$ | $(V-K)_{RGB}$ | $(B-V)$ | $(B-V)_{AGB+RGB}$ |
|---|---|---|---|---|---|---|---|---|
| NGC 1756 | 0.62 | 0.54 | 0.48 | – | – | – | 0.40 | 0.21 |
| NGC 1783 | 1.02 | 0.74 | 0.60 | – | – | – | – | – |
| NGC 1806 | 1.07 | 0.75 | 0.51 | – | – | – | – | – |
| NGC 1831 | 1.33 | 0.67 | 0.51 | 1.40 | 0.50 | 0.38 | 0.34 | 0.29 |
| NGC 1868 | 0.76 | 0.48 | 0.41 | – | – | – | 0.45 | 0.23 |
| NGC 1978 | 1.14 | 0.73 | 0.58 | – | – | – | – | – |
| NGC 1987 | 1.22 | 0.76 | 0.53 | – | – | – | 0.52 | 0.40 |
| NGC 2107 | 1.09 | 1.02 | 1.12 | – | – | – | 0.38 | 0.37 |
| NGC 2108 | 1.24 | 1.04 | 1.10 | 2.54 | 2.20 | 1.96 | 0.58 | 0.39 |
| NGC 2162 | 0.89 | 0.72 | 0.68 | 2.12 | 0.58 | -0.76 | 0.68 | 0.51 |
| NGC 2173 | 1.14 | 0.87 | 0.56 | 2.90 | 2.02 | 1.46 | 0.84 | 0.80 |
| NGC 2209 | 1.58 | 0.80 | 0.80 | 2.92 | 0.40 | 0.40 | 0.53 | 0.18 |